\documentclass[12pt]{iopart}
\usepackage{tabulary,graphicx}
\usepackage{autonum}
\begin{document}
\title[]{Rare decays of the $B_s$ -- meson into four charged leptons in the framework of the Standard Model}

\author{A. V. Danilina$^{1,2}$, N. V. Nikitin$^{2,3,4}$\footnote{Present address:
Federal State Budget Educational Institution of Higher Education M.V.Lomonosov Moscow State University, Skobeltsyn Institute of Nuclear Physics (SINP MSU), 1(2), Leninskie gory, GSP-1, Moscow 119991, Russian Federation}}

\address{$^1$Skobeltsyn Institute of Nuclear Physics, Moscow, Russia }
\address{$^2$NRC "Kurchatov Institute" – ITEP, Moscow, Russia}
\address{$^3$Lomonosov Moscow State University, Physics Faculty, Moscow, Russia}
\address{$^4$The Moscow Institute of Physics and Technology, Dolgoprudny, Moscow Region, Russia}

\ead{anna.danilina@cern.ch, Nikolai.Nikitine@cern.ch}

\begin{abstract}
In the framework of the Standard Model we present new theoretical predictions for the branching ratios, double and single differential distributions and forward -- backward leptonic asymmetries for the $\bar B_s \to \mu^+ \mu^- e^+ e^-$ decay. In our consideration we take into account  the $\phi(1020)$ -- resonance contribution; the main contributions of four charmonium resonances : $\psi(3770)$, $\psi(4040)$, $\psi(4160)$ and $\psi(4415)$; $u\bar{u}$ -- resonant contribution from $\rho(770)$ and $\omega(782)$; “tails” contributions from $J/\psi$ and $\psi(2S)$ resonances; non -- resonant contribution of the $b\,\bar{b}$ - pairs, bremsstrahlung and the  contribution of the weak annihilation. We provide the prediction for the branching ratio of $\bar B_s \to \mu^+ \mu^- e^+ e^-$ decay with and without the $\phi(1020)$ -- resonance contribution.  We use the model of vector meson dominance (VMD) for calculation of resonances contributions and take into account all substantive terms in $\bar B_s \to \mu^+ \mu^- e^+ e^-$ amplitude that was not considered in the previously papers.
\end{abstract}

\section{Introduction}
The search for some deviations from the Standard Model (SM) predictions is the great challenge of high energy physics at the current moment. However, many experimental results are unclear and not explained in the SM. For example, an indication of the existence of dark matter in astrophysical experiments was well established~\cite{DarkMatter}. At the present days it is the most powerful argument for the existence of new physics. There are a number of experimental indications of deviations from the SM predictions, such as violation of lepton universality in rare semileptonic $B$ -- decays; data on the anomalous magnetic moment of the muon, diverging from the SM predictions~\cite{AnMu}; hints of a discrepancy between the experimental and predicted value of the partial widths for the $B_s \to \mu^+\mu^-$ decays~\cite{CMS:2014xfa,Aaboud:2016ire,Aaij:2017vad,Aaij:2021}.

It is shows the great importance of the searching for statistically significant effects beyond
the Standard Model (BSM) physics at modern and future accelerators, where, in contrast to astrophysical observations, the experiment is completely controlled and the systematic uncertainties are under the supervision.

In recent years, experimental data on weak decays of $b$ -- and $c$ -- quarks have gained importance in the study of BSM effects. Recent anomalies found in weak $B$ -- meson decays may be crucial~\cite{Anom}. There have been suggestions that a breakthrough in high energy physics can occur precisely from the analysis of weak decays of $b$ -- and $c$ -- quarks. In the next two years, for the implementation of this idea, a large amount of new data in heavy flavour physics is expected, primarily from the BELLE -- II (KEK) and LHCb (LHC, CERN) experiments~\cite{Belle}. It will enable to advance in the study of very rare four -- leptonic decays of $B$ -- mesons. These decays are suppressed in the SM, which makes them effective tools for searching for the effects of new physics. In that perspective, it is necessary to obtain reliable theoretical predictions for such extremely $B$ -- mesons decays and to understand  which decays characteristics are the best for the looking for the manifestation of BSM effects.\\
The rare four -- leptonic decays of B -- mesons in SM can be divided into two types. The decays of the first type include a large number of electromagnetic and weak processes at the tree level realizing the final lepton state. A typical example of such decays are the decay $ B^- \to \mu^+ \mu^- \mu^+ \bar\nu_{\mu}$ and similar decays of the charged B mesons. Decays of the second type are forbidden at the tree level and occur in higher orders of perturbation theory due to loop diagrams of the “penguin” and/or “box” type. The contribution of such processes can be described using Flavor Changing  Neutral Currents (FCNC). In SM they are due to loop quantum diagramms, where the contribution of new virtual particles can be noticeable and measurable. An example of the second decays type is the processes $\bar B_{d,s} \to \mu^+ \mu^- e^+ e^- $ and $\bar B_{d,s} \to \mu^+ \mu^- \mu^+ \mu^-$.  The first and second types of processes are being studied at the LHC and are planned to be studied at the Belle II facility. Currently, only upper limits are found for the partial widths of four -- leptonic decays $B_d$ and $B_s$ ~\cite{Aaij:2016kfs},~\cite{Aaij:2013lla},~\cite{4mu2021}.
Thus, a theoretical study of the rare leptonic decays of neutral $B^0_{d,s}$ -- mesons is one of great problem for further investigations on the LHC and other experiments. At the moment, for these decays  there is only a prediction given in the ~\cite{Dincer:2003zq}, which does not take into account the many resonant contributions, and an rough estimation from~\cite{Danilina:2018uzr}.

In this paper we present the detailed calculation of the branching ratio of
$\bar B_s \to \mu^+ \mu^- e^+ e^- $ and differential characteristics, taking into account contributions of $\phi(1020)$ -- resonance, the main contributions of charmonium resonances, non -- resonance contribution of the $b\,\bar{b}$ -- pairs, leading contribution of weak annihilation and bremsstrahlung. 
\section{Effective Hamiltonian}
\label{sec:B2lllnuHeff}
The FCNC $b \to q$ (where q = \{d,s\}) effective Hamiltonian have the form of the Wilson expansion~\cite{BrMu}:
$$
{\cal H}_{eff}^{b \to q}(x,\mu)\, =\,  \frac{G_F}{\sqrt{2}}\,V^*_{tq}\, V_{tb}\,\sum\limits_i\, C_i(\mu)\,O^{b \to q}_i(x),
$$
where $\mu \approx m_b$ is the scale parameter which separats short and long distance contributions of the strong interactions, $G_F$ is the Fermi constant, $V_{tq}$ and  $V_{tb}$ are the matrix elements of the Cabibbo-Kobayashi-Maskawa (CKM) matrix.  The light degrees of freedom of the SM, such as $u, d, s, c,$ and $b$ - quarks, leptons, photons and gluons are contained in the basis operators $O^{b \to q}_i(x,\mu)$. The heavy degrees
of freedom, $W, Z,$ and $t$ - quark, are introduced into the Wilson coefficients
$C_i(\mu)$. The sign of the Wilson coefficients is determined by the condition $C_2(M_W) = -1$.

To calculate the amplitude of the 
decay $\bar B_s \to \mu^+ \mu^- e^+ e^-$ the Hamiltonian for the FCNC transition $b\to s\ell^+\ell^-$ is used: 

\begin{eqnarray}\label{eq:b2sll}
&&{ {\cal H}_{eff}^{b\to s\ell^{+}\ell^{-}}(x,\mu)\, =\, 
{\frac{G_{F}}{\sqrt2}}\, {\frac{\alpha_{em}}{2\pi}}\, V_{tb}V^*_{ts}\, 
\left[
\,-2\, {\frac{C_{7\gamma}(\mu)}{q^2}}\, 
        \Big\{ m_b\,
 \left (\bar s\, i\sigma_{\mu\nu}\left (1+\gamma_5\right )q^{\nu}b\right )
        \nonumber \right .}\\
&&\qquad\qquad\quad {+\, m_s\, 
 \left (\bar s\, i\sigma_{\mu\nu}\left (1-\gamma_5\right )q^{\nu}b\right )
        \Big \}
\cdot\left ({\bar \ell}\gamma^{\mu}\ell \right )}\\
&&\qquad\qquad\quad{+\, 
{C_{9V}^{eff}(\mu, q^2)}\left (\bar s  O_{\mu} b \right )
\cdot\left ({\bar \ell}\gamma^{\mu}\ell \right )\, +\, 
C_{10A}(\mu)\left (\bar s  O_{\mu} b \right )
\cdot\left ({\bar \ell}\gamma^{\mu}\gamma_{5}\ell \right ) \Big]},\nonumber 
\end{eqnarray}
 where ${O_{\mu} = \gamma_\mu (I - \gamma^5)}$ and ${q^\nu}$ is the four - momentum of ${\ell^+\ell^-}$ --  pair, the matrix $\gamma^5$ is
defined as $\gamma^5 = i \gamma^0 \gamma^1 \gamma^2 \gamma^3$, $\varepsilon^{0123} = -1$ and $\sigma_{\mu\,\nu} = \frac{i}{2}[\gamma_\mu,\gamma_\nu]$.

In accordance with definition (Eq.\ref{eq:b2sll}) the Hamiltonian of the electromagnetic interaction has the form:

\begin{eqnarray}\label{eq:belm}
{\cal H}_{em}(x)\, =\, -\, e\,\sum\limits_f\, Q_f
\left (
\bar f(x)\,\gamma^{\mu} f(x)
\right )\,
A_{\mu}(x)\, =\, -\, j_{em}^{\mu}\,A_{\mu}(x),
\end{eqnarray}
where the charge $e = |e| > 0$ is normalized by $e^2 = 4 \pi
\alpha_{em}$; $\alpha_{em} \approx 1/137$ is the fine structure constant;
$Q_f$ is the charge of the fermion of flavor $f$ in units of the
$e$, $f(x)$ is the fermionic field of flavor $f$ and $A_{\mu}
(x)$ is the four-potential of the electromagnetic field.
\\
The Hamiltonian for the weak annihilation has the form 
\begin{eqnarray}
\label{WEAK}
{\cal H}_{eff}^{\bar{B_q}-Q\bar{Q}}(x)\, =\,- 
{\frac{G_{F}}{\sqrt2}}\,  V_{Qb}V^*_{Qq}\,a_1(\mu)(\bar{q}\,O^\mu\,b)(\bar{Q}\,O_\mu\,Q),
\end{eqnarray}
where $Q =\{u,c\}$ and $a_1(\mu = \textrm{5 GeV}) \approx - 0.13$.

\section{Main contributions to the $\bar B_s \to \mu^+ \mu^- e^+ e^- $ decay}
\label{s3}
\begin{figure}[h!]
\begin{tabular}{c}
\includegraphics[width=0.55\linewidth]{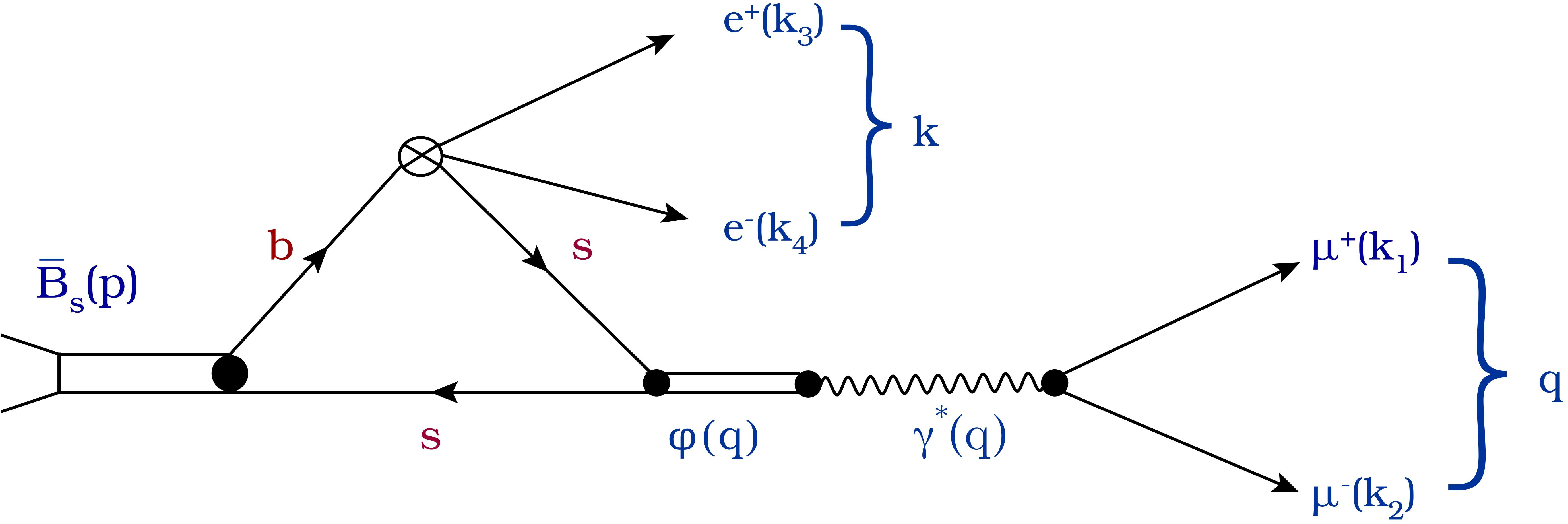}
\includegraphics[width=0.55\linewidth]{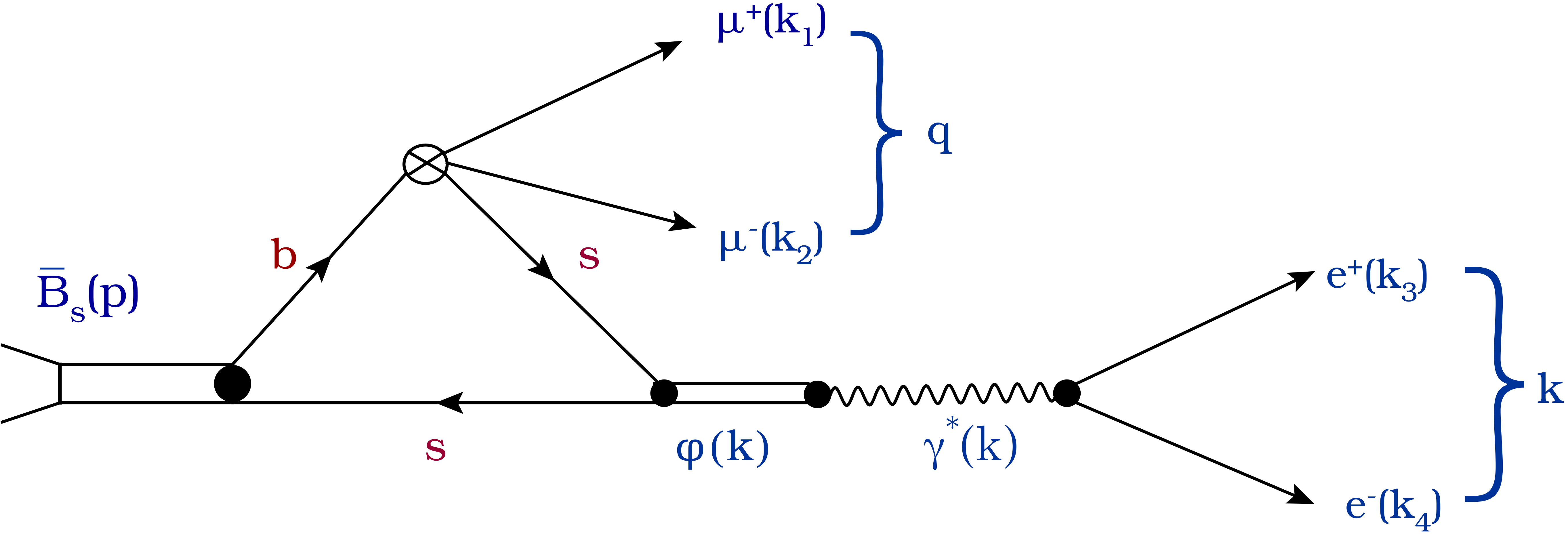} 
\end{tabular}
\caption{\label{fig:Fphi} 
Emission diagram of a virtual photon by a $s$ -- quark of $B_s$ meson.}
\end{figure}

\begin{figure}[h!]
\begin{tabular}{c}
\includegraphics[width=0.55\linewidth]{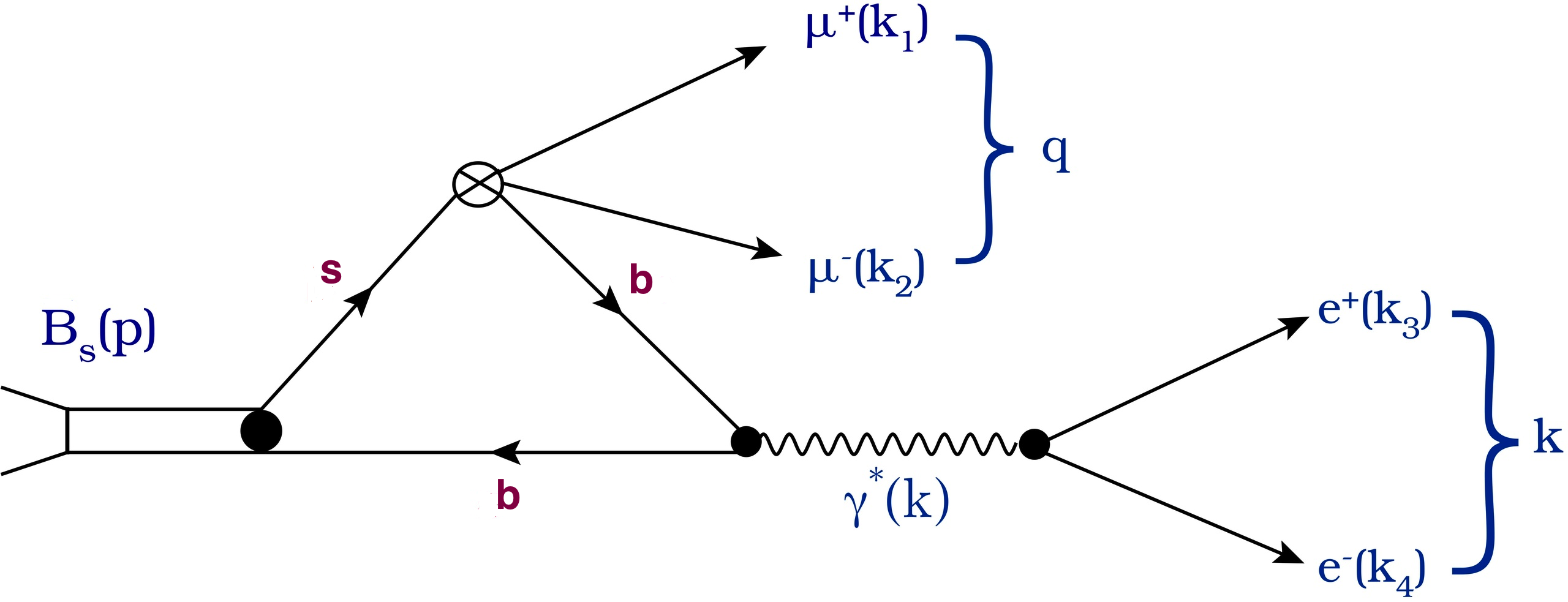}\,
\includegraphics[width=0.55\linewidth]{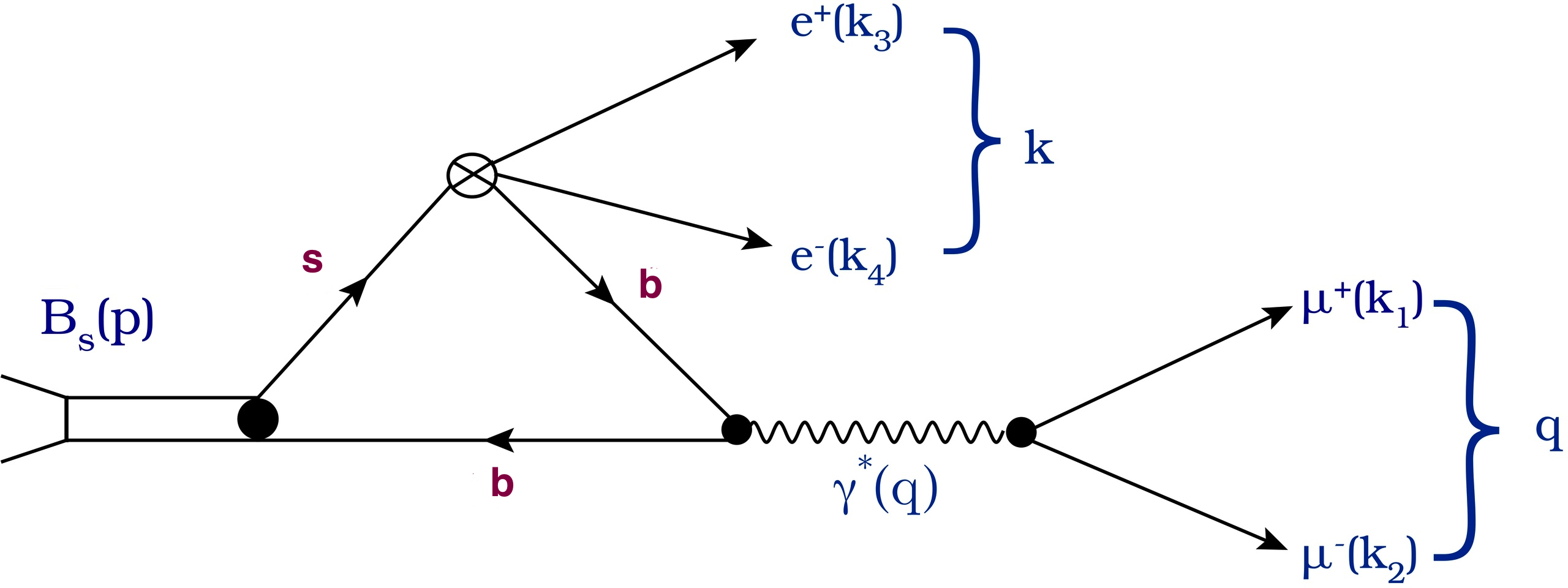} 
\end{tabular}
\caption{\label{fig:Fbbarb} 
Emission diagrams of a virtual photon by a $b$ -- quark of $B_s$ meson}
\end{figure}

\begin{figure}[h!]
\begin{tabular}{cc}
\includegraphics[width=0.35\linewidth]{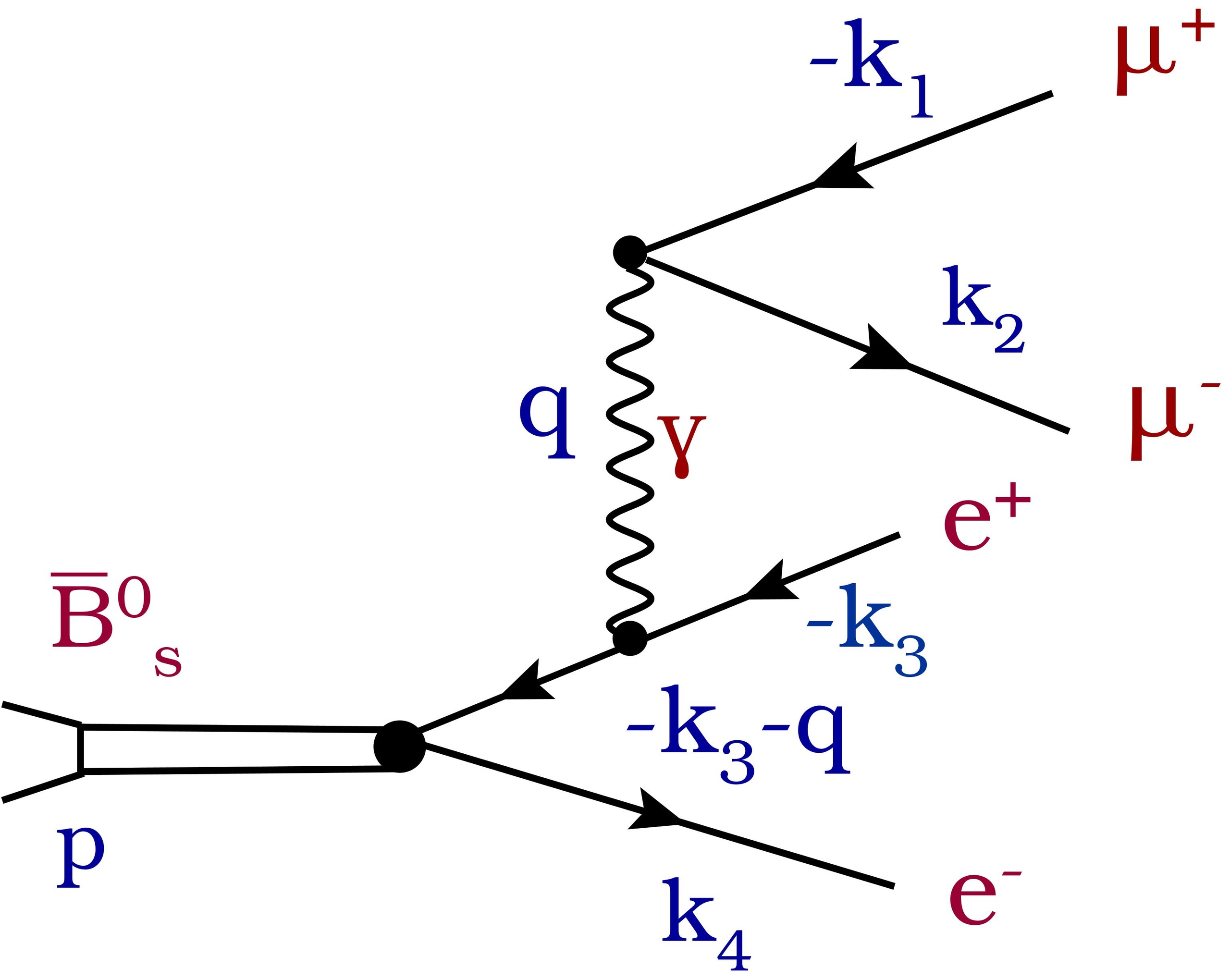} &\,\,\,\,\,\,\,\,\,\,\,\,\,\,\,\,\,\,\,\,\,\,\,\,\,\,\,\,\,\,\,\,\,\,\,\,\,\,\,\,\,\,\,\,\,
\includegraphics[width=0.35\linewidth]{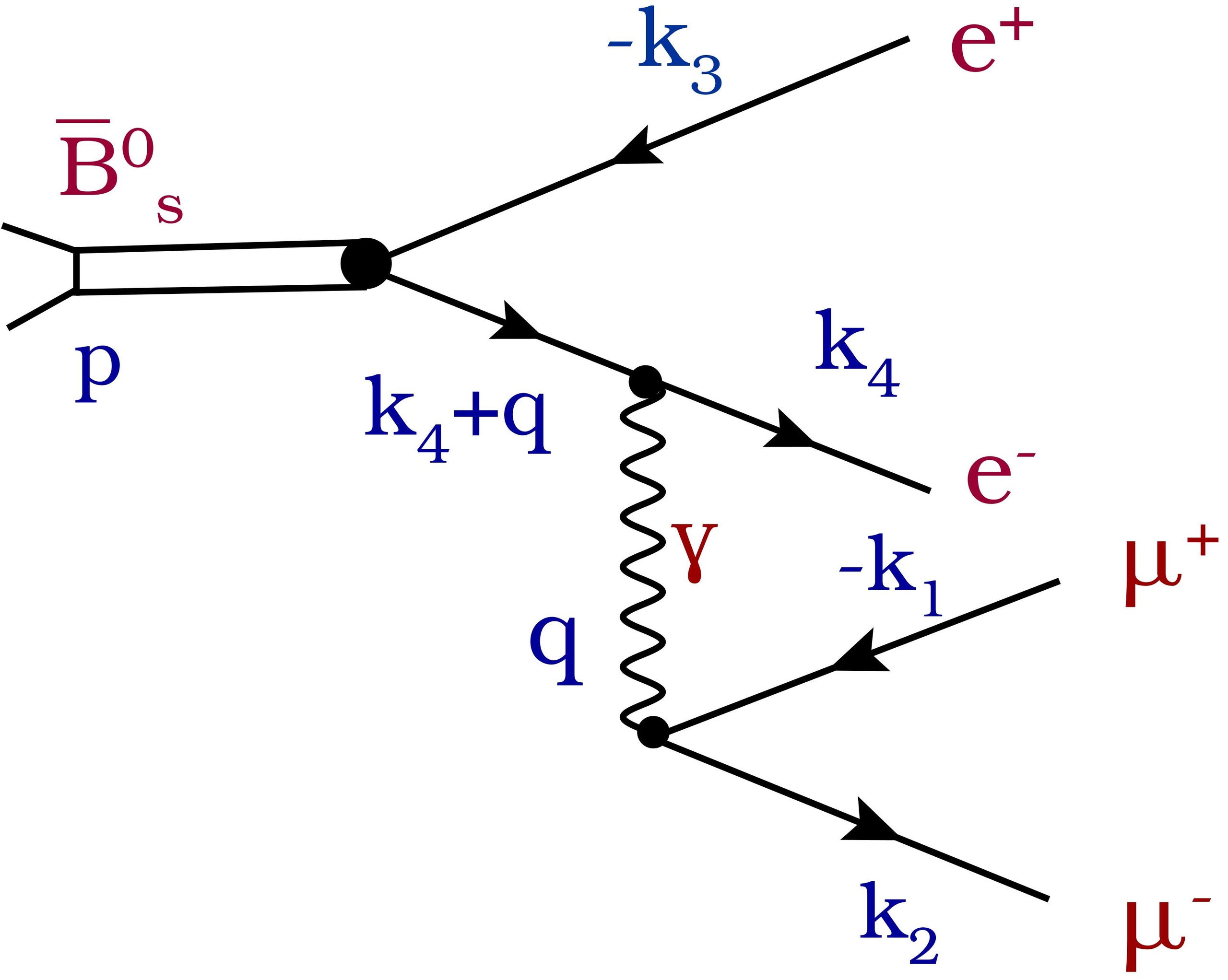} \\
\includegraphics[width=0.35\linewidth]{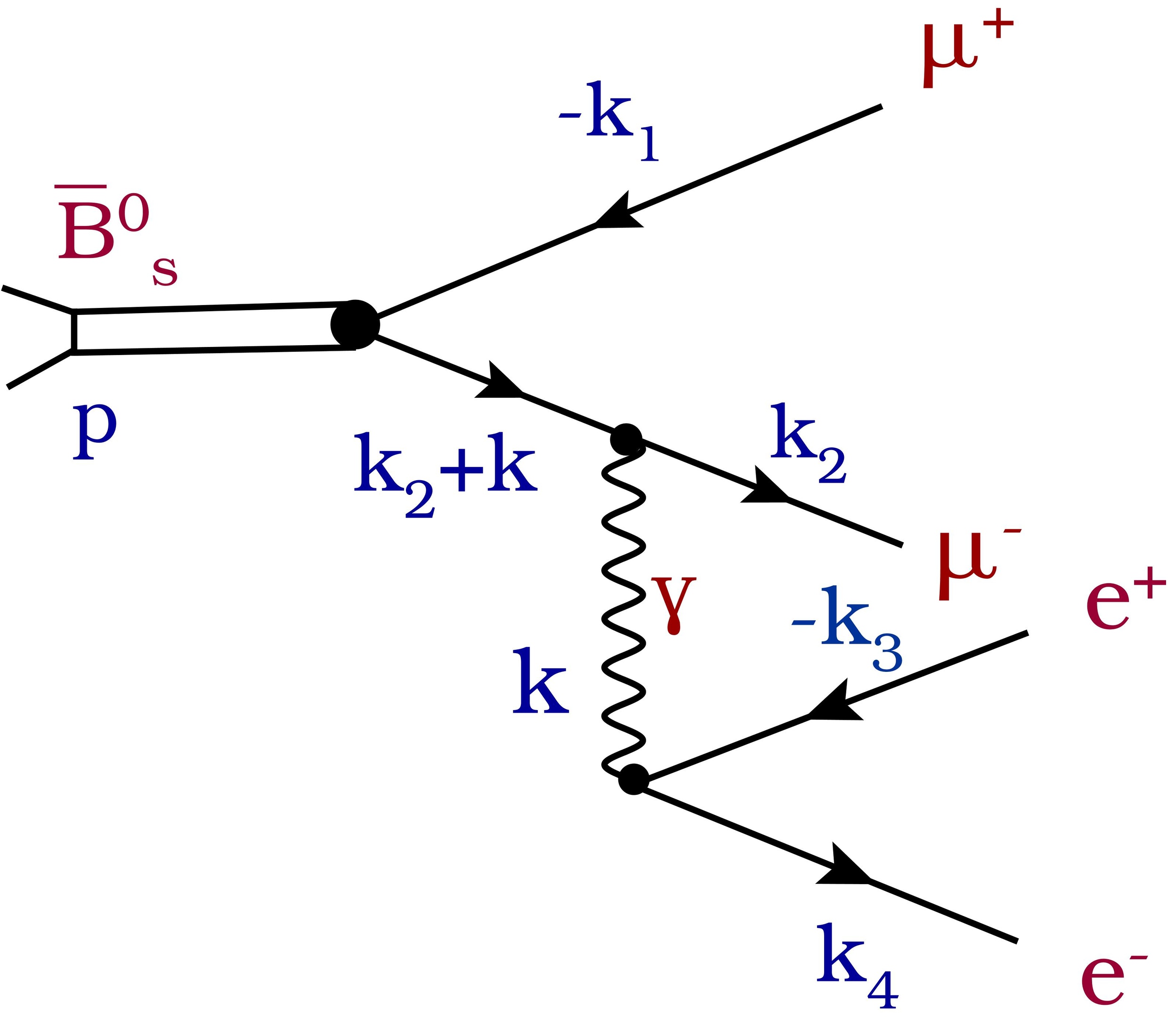} &\,\,\,\,\,\,\,\,\,\,\,\,\,\,\,\,\,\,\,\,\,\,\,\,\,\,\,\,\,\,\,\,\,\,\,\,\,\,\,\,\,\,\,\,\,
\includegraphics[width=0.35\linewidth]{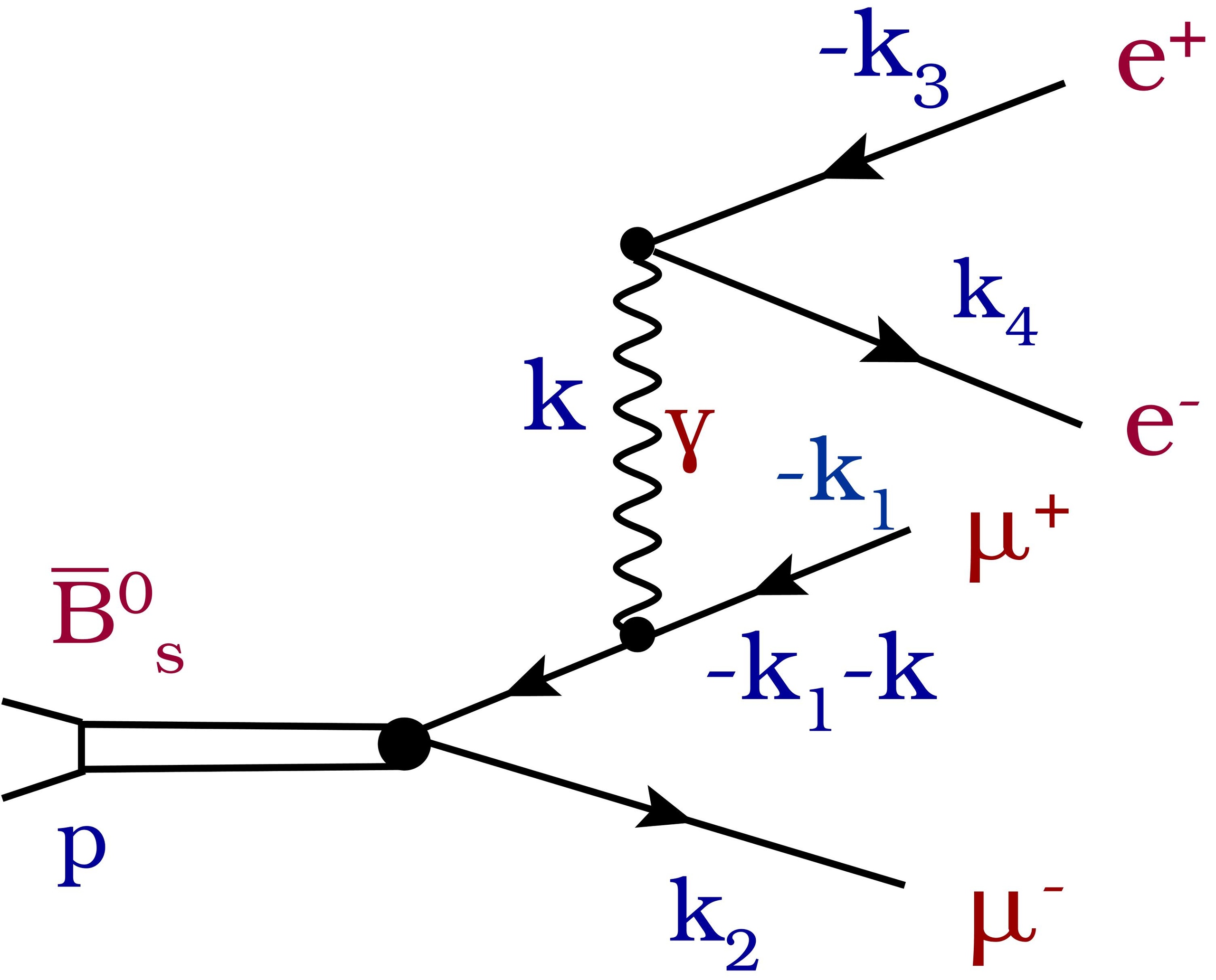} \\
\end{tabular}
\caption{\label{fig:Fbrem} 
Emission diagram of a virtual photon by leptons in the final state.}
\end{figure}

There are six main types of contributions are required for the description of the decays 
$\bar B_s\to\mu^+(k_1)\mu^-(k_2)e^+(k_3)e^-(k_4)$. The first type arises in the situation when a virtual
photon is emitted from $s$ -- quark (see Fig.~\ref{fig:Fphi}). The second type is related to bremsstrahlung, when a
virtual photon is emitted by the lepton in the final state (see Fig.~\ref{fig:Fbrem}). The third and fourth types reflect the contributions from $u\bar{u}$ and $c\bar{c}$. The fifth type  corresponding to the $b\bar{b}$ - pair's contribution shown on the Fig.~\ref{fig:Fbbarb}. And the last one tied to the weak annihilation processes (see Fig.~\ref{fig:WA}). 
The momenta $q = k_1 + k_2$ and $k = k_3 + k_4$ (see
Appendix~\ref{fig:cinematic}).

Using the Hamiltonians (Eq.\ref{eq:b2sll},\ref{eq:belm} and  \ref{WEAK}) we can find, that the amplitude for a process of virtual photon emission by $s$ -- quark as well as the $u\bar{u}$, $c\bar{c}$ and $b\bar{b}$ - pairs contributions can be presented as

\begin{eqnarray}
\label{Amp}
\fl
{\cal M}_{fi}^{(1234)} = i\,\sqrt2\,G_{F}\,\alpha_{em}^2\, V_{tb}V^*_{ts}M_1\Bigg[\nonumber\\
 \hphantom{+}\fl\Bigg[-\frac{a^{(VV)}}{M^2_1}\,\varepsilon_{\mu\,\alpha\, k\, q}\, -\, i b^{(VV)}\, g_{\mu \alpha}\, +\, 2i\,\frac{c^{(VV)}}{M^2_1}\, q_{\alpha} k_{\mu}\,\Bigg]j^{\mu} (k_2,\, k_1)\, J^{\alpha} (k_4,\, k_3)\, +\nonumber\\
 \fl+\Bigg[-\frac{a^{(VA)}}{M^2_1}\,\varepsilon_{\mu\,\alpha\, k\, q}\, -\, i  b^{(VA)}\, g_{\mu \alpha}\, +\, 2i\,\frac{c^{(VA)}}{M^2_1}\, q_{\alpha} k_{\mu}\, +\,i\,\frac{g^{(VA)}}{M^2_1}\, k_{\mu} k_{\alpha}\,\Bigg]j^{\mu} (k_2,\, k_1)\, J^{\alpha\,5} (k_4,\, k_3)\,+\\
 \fl+\Bigg[-\frac{a^{(AV)}}{M^2_1}\,\varepsilon_{\mu\,\alpha\, k\, q}\, -\, i  b^{(AV)}\, g_{\mu \alpha}\, +\, 2i\,\frac{c^{(AV)}}{M^2_1}\, q_{\alpha} k_{\mu}\,+\,i\,\frac{d^{(AV)}}{M^2_1}\, q_{\mu} k_{\alpha}\,\Bigg]j^{\mu\,5} (k_2,\, k_1)\, J^{\alpha} (k_4,\, k_3)\,+\nonumber\\
 \fl+\Bigg[-\frac{a^{(AA)}}{M^2_1}\,\varepsilon_{\mu\,\alpha\, k\, q}\, -\, i  b^{(AA)}\, g_{\mu \alpha}\, +\, 2i\,\frac{c^{(AA)}}{M^2_1}\, q_{\alpha} k_{\mu}\,+\,i\,\frac{d^{(AA)}}{M^2_1}\, q_{\mu} k_{\alpha}\,+\,i\,\frac{g^{(AA)}}{M^2_1}\, k_{\mu} k_{\alpha}\,\Bigg]j^{\mu\,5} (k_2,\, k_1)\, J^{\alpha\,5} (k_4,\, k_3)\,\Bigg],\nonumber
\end{eqnarray}
where currents are defined as 
\begin{eqnarray}
j^{\mu} (k_2,\, k_1) = \bar\mu(k_2)\,\gamma^\mu\mu(-k_1),\,\, J^{\alpha} (k_4,\, k_3) = \bar e(k_4)\,\gamma^\alpha\,e(-k_3);\nonumber\\
j^{\mu\,5} (k_2,\, k_1) = \bar\mu(k_2)\,\gamma^\mu\,\gamma^5\,\mu(-k_1),\,\, J^{\alpha\,5} (k_4,\, k_3) = \bar e(k_4)\,\gamma^\alpha\,\gamma^5\,e(-k_3).\nonumber 
\end{eqnarray}

Dimensionless functions
$a^{(IJ)} \equiv a^{(IJ)} (x_{12},\, x_{34})$, $b^{(IJ)} \equiv b^{(IJ)} (x_{12},\, x_{34})$, $c^{(IJ)} \equiv c^{(IJ)} (x_{12},\, x_{34}),$\, $d^{(IJ)} \equiv d^{(IJ)} (x_{12},\, x_{34})$ and $g^{(IJ)} \equiv g^{(IJ)} (x_{12},\, x_{34})$, where ${IJ\, =\,\{VV,\, VA,\, AV,\, AA\}}$ are defined in~\ref{sec;ABC}.

The contribution to the decay amplitude (Eq.\ref{Amp}) from Fig.\ref{fig:Fphi} may be calculated using the Vector Meson
Dominance model (VMD). Here we consider two cases: a virtual photon emission produces a $\mu^+\mu^-$ -- pair (left diagram) and a virtual photon emission produce a $e^+ e^-$ -- pair (right diagram). In both cases there is the intermediate vector ${\phi (1020)}$ -- meson that we take into account in the framework of VMD (as reflected in $a^{(IJ)}$, $b^{(IJ)}$, $c^{(IJ)}$, $d^{(IJ)}$ and $g^{(IJ)}$ coefficients in~\ref{sec;ABC}). As the further calculations will show, the ${\phi (1020)}$ -- resonance gives a leading contribution to the amplitude of $\bar B_s \to \mu^+ \mu^- e^+ e^- $ decay.

Charmonium vector resonances and $\rho^0(770)$, $\omega(782)$  contributions are arised from the effective Hamiltonian for the transition ${b \to s\ell^{+}\ell^{-}}$ (Eq.\ref{eq:b2sll}) and contained in coefficients $a^{(IJ)}$, $b^{(IJ)}$ and  $c^{(IJ)}$. The resonant contributions from $J/\psi$, $\psi(2S)$ ..., $\rho^0(770)$ and $\omega(782)$  are contained in the coefficient $C^{eff}_{9V}(\mu, q^2)$. This coefficient consists of the fixed part, depending on $\mu$ -- scale, $c\bar c\,\,\textrm{and}\, u\bar u\,$ quark loops contribution and vector resonances contribution. \\
The structure of the $C^{eff}_{9V}(\mu, q^2)$ may be
presented as
$$
{C^{eff}_{9V}(\mu, q^2) \, =\,
C_{9V}(\mu)\, +}\,{\Delta C_{9V}^{c\bar c\, +\, u\bar u}(\mu, s)},  
\nonumber
$$
where $C_{9V}(\mu)$ is the Wilson coefficient ($C_{9V}(\mu=m_b)=-4.21 $) and $\Delta C_{9V}^{c\bar c\, +\, u\bar u}(\mu, s)$ is non-perturbative correction, consisting of loop and resonant effects.
\\ In the factorization approximation the structure of the $\Delta C_{9V}^{c\bar c\,}(\mu, s)$ is given in Ref. \cite{MNS}.

In according to the experimental procedure of the $J/\psi$ and $\psi(2S)$ contributions exclusion \cite{Aaij:2016kfs,Aaij:2013lla,4mu2021} in the theoretical calculation of the differential characteristics and branching ratio of $\bar B_s \to \mu^+ \mu^- e^+ e^- $ decay we leave only tails from $J/\psi$ and $\psi(2S)$  resonances.  The  higher vector charmonium contributions are not cut, since they don't  overlap less significant contributions.

The next contribution to the (Eq.~\ref{Amp}) amplitude is the non-resonant contribution from $b\bar b$ -- pairs (Fig.\ref{fig:Fbbarb}).

Using the numerical calculation ${F_i(0,\, 0)}$  and ${M_{R_i}}$ from Ref. \cite{MNK}, it is possible to find the following parameterization for the non-resonant form factor:
$$
{F_i (q_1^2,\, q_2^2)}\, {=\,\frac{F_i (q_1^2 = 0,\, q_2^2= 0)}{\displaystyle\left ( 1\, -\,\frac{q_1^2}{{M_{R_i}^2}}\right )\left ( 1\, -\,\frac{q_2^2}{{M_{\Upsilon(1S)}^2}}\right )\,}},
$$
where ${i\, =\,\{V,\, A,\, TV,\, TA\}}$.

As $M_{B^*_s} > M_1$, this pole lies outside of the kinematically allowed range of the decay
$\bar B_s \to \mu^+ \mu^- e^+ e^- $.  The existence of this pole is taken into account when choosing the pole parameterisation of the form factors.

The next set of diagrams is related to bremsstrahlung, when a virtual photon is emitted by one of lepton in the final state (see Fig. \ref{fig:Fbrem}). There are four types of diagrams, for the photon emission by each lepton in the final state accordingly.

The amplitude for the ${\mu^+\mu^-}$-- pair emitted by electron and positron in the final state is:
\begin{eqnarray}
\label{MuBrem}
{{\cal M}_{fi}^{(\mu)}} &{=}&
i{\sqrt{2}\, G_{\textrm{F}}\,\alpha^2_{em}\, V_{tb}\, V_{ts}^*\, \Big (\overline{\mu}(k_2){\gamma^\mu}\mu(-k_1) \Big )\,
\Big [}
\nonumber\\
&&{ i\,{ d^{(VP)}} (x_{12},\, x_{123},\, x_{124})\, k_{\mu}\,
\Big (\overline{e}(k_4){\gamma^5} e(-k_3) \Big )\, +}
\\
&{+}&
{ f^{(VT)}} {(x_{12},\, x_{123},\, x_{124})\,\varepsilon_{\mu \nu \alpha \beta}\, p^{\nu}\,
\Big (\overline{e}(k_4){\gamma^\alpha \gamma^\beta} e(-k_3) \Big )
\Big]}.
\nonumber
\end{eqnarray}
The similar expression can be obtained for the $e^+e^-$ -- pair emitted by muons the in final state. But for short, we do not write it here.
For the calculation of the pole structure in bremsstrahlung non-zero lepton masses should be taken into account. It is done in the (Eq.\ref{MuBrem}) and in the expression for four-particle phase space in the~\ref{sec;kinemat4}.
\\
The last contribution that we consider is the weak annihilation processes. These ones arise from lowest -- order diagrams describing contribution of $c\bar{c}$ and $u\bar{u}$ the after integrating out the $W$ -- bosons degrees of freedom. We taken into consideration axial anomaly contribution gives the following amplitude structure:

\begin{figure}[h!]
\begin{center}
\begin{tabular}{c}
\includegraphics[width=0.59\linewidth]{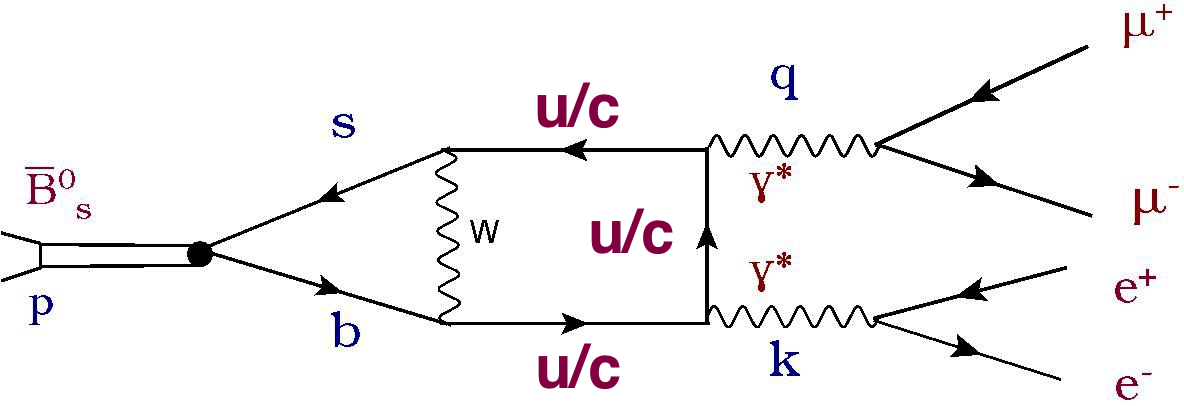}\\
\includegraphics[width=0.59\linewidth]{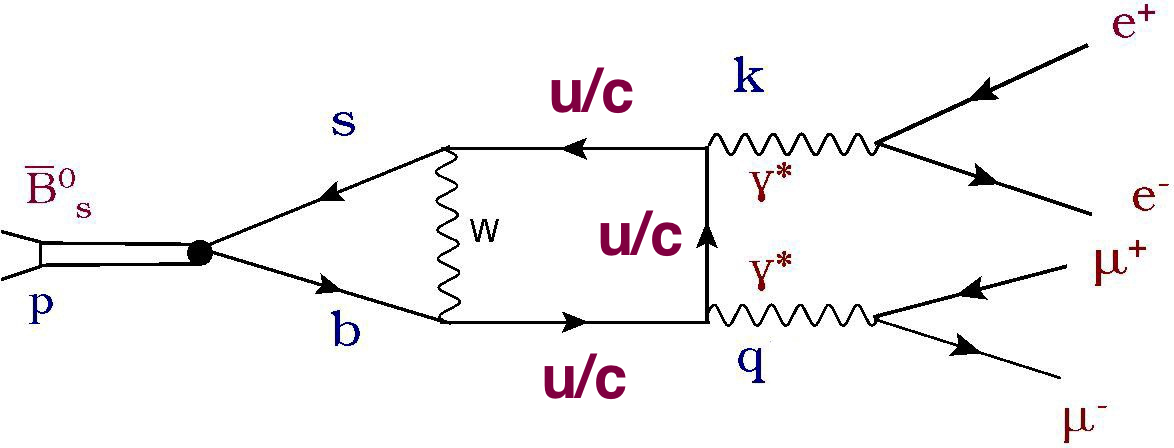} 
\end{tabular}
\caption{
The diagrams for the weak annihilation processes.}
\label{fig:WA}
\end{center}
\end{figure}

\begin{eqnarray}
\
&&{{\cal M}_{fi}^{(WA)}}\,{= \,i\frac{32 \sqrt{2}}{3 \pi}\,\frac{G_{\textrm{F}}}{M^3_1}\,\alpha^2_{em}\, \left ( V_{ub}\, V_{us}^*+V_{cb}\, V_{cs}^* \right )\, a_1 (\mu)\, {\hat f}_{B_s}} 
\nonumber \\
&&
{\frac{1}{x_{12}\, x_{34}}\,\varepsilon_{\mu\alpha k q}
\Big (\overline{\mu}(k_2)\,{\gamma^\mu}\mu(-k_1) \Big )\,
\Big (\overline{e}(k_4)\,{\gamma^\alpha} e(-k_3) \Big )},
\nonumber
\end{eqnarray}
where ${\hat f}_{B_s} = \frac{f_{B_s}}{M_1}$. Here we neglect ${\left (m_c/M_1\right )^2}$ and ${\left (m_u/M_1\right )^2}$ corrections.
\\
The weak annihilation contribution is enhanced at small four–momenta, but even here
it is suppressed by a power of a heavy quark mass
compared to the contributions discussed in the previous paragraphs.

\section{The branching ratio calculation}

The extremely rare $B_{d,s}$ -- decays provide the vigorous studies in the direction on the searches of physics beyond the Standard Model. From this point of view, detailed investigation of $\bar B_s \to \mu^+ \mu^- e^+ e^- $ decays is
very promising. It demands the correct simulation of this decays. For that reason we produce the Monte Carlo model for the EvtGen package~\cite{Evt}. This model provides correct simulation for $\bar B_s \to \mu^+ \mu^- e^+ e^- $ decay and include all theoretical contributions which described in the Sec.~\ref{s3}.
The tools of the EvtGen package allows to calculate the numerical value of the branching ratio of the $\bar B_s \to \mu^+ \mu^- e^+ e^- $ decay.

The total amplitude of the decay $\bar B_s \to \mu^+ \mu^- e^+ e^- $  can be presented in the form:
\begin{eqnarray}
\label{Mtot}
{{\cal M}^{(tot)}} = {\cal M}_{fi}^{(BS\,\, e + \mu)} + {\cal M}_{fi}^{(\phi)} + {\cal M}_{fi}^{(b \bar b)} +{\cal M}_{fi}^{(c \bar c + u \bar u)} + {\cal M}_{fi}^{(WA)} = \nonumber\\ \,\,\,\,\,\,\,\,\,\,\,\,\,\,\,\,\,\,\,\,\,\,\,\,\,\,\,\,\,\, \sqrt{2}\, G_{F}\,\alpha^2_{em}\, V_{tb}\, V_{ts}^*\,\sum\limits_{L}L\,j_1\,j_2,
\nonumber
\end{eqnarray}
where $j_1$ and $j_2$ are lepton currents, $L$ are the Lorentz structures contributed to this matrix element.\\
The differential branching ratio of the decay $\bar B_s \to \mu^+ \mu^- e^+ e^- $ has the form
\begin{eqnarray}
\label{dBr1234common}
d \textrm{Br} \left (\bar B_s \to \mu^+ \mu^- e^+ e^- \right ) \, =\, \tau_{B_s}\,
\frac{\sum\limits_{s_1,\, s_2,\, s_3,\, s_4}\left | {\cal M}_{fi}^{(tot)} \right |^2}{2 M_1}\, d\Phi_4 = \nonumber\\
\,\,\,\,\,\,\,\,\,\,\,\,\,\,\,\,\,\,\,\,\,\,\,\,\,\,\,\,\,\,
= \frac{G^2_{F}\,\alpha^4_{em}\, |V_{tb}\, V_{ts}^*|^2}{M_1}\,\tau_{B_s}\,\left |\sum\limits_{L}L\,j_1\,j_2\right |^2\,d\Phi_4
\nonumber
\end{eqnarray}
where $\tau_{B_s}$ is the lifetime of the $B_s$--meson, four-particle
phase space $d\Phi_4$ is defined by (Eq.\ref{dPhi1234}). The summation is performed over the spins of the final leptons $s_1, s_2, s_3$ and $s_4$.  Full integration may be performed only
numerically.
By implementing built-in opportunity of the EvtGen package we realized the multidimensional integrator based on the effective geometry Monte Carlo algorithm. Within this approach, the branching ratio of the $\bar B_s \to \mu^+ \mu^- e^+ e^- $ decay may calculate as

\begin{eqnarray}
\label{BrfromEvt}
{\textrm{Br}\,\left (\bar B_s \to \mu^+ \mu^- e^+ e^- \right )\,\approx\,
\frac{\alpha^4_{em}\, \left |V_{tb}\, V_{ts}^*\right |^2}{3\cdot 2^{13}\cdot \pi^5}\,
\tau_{B_s}\, G^2_{F}\, M^5_1\,\frac{N_0}{N_{{tot}}}\,|X|^2},\nonumber
\end{eqnarray}
where ${N_{\textrm{tot}}}$ is total and ${N_0}$ is accepted numbers of events produced by EvtGen. $|X|^2 \equiv \max\frac{|\sum\limits_{L}L\,j_1\,j_2\,|^2}{M^2_1}$ is the dimensionless maximum of the decay matrix element. In the last formula we used the massless approach of the phase space. In the framework of the EvtGen we have the four -- particle phase space with non -- zero leptonic masses. The approach of multidimensional integration in the EvtGen has been tested on the already known branching ratios of several decays, as well as on the various functions, including the singular. 
Numerical integration for the $\bar B_s \to \mu^+ \mu^- e^+ e^-$ decay gives:
$$
{\textrm{Br}\,\left (\bar B_s \to \mu^+ \mu^- e^+ e^- \right ) \approx (61\pm 12)*10^{-10}}.
$$
Here the $J/\psi$ and $\psi (2S)$ resonances contributions are excluded from the calculation of the $\textrm{Br}\,\left (\bar B_s \to \mu^+ \mu^- e^+ e^- \right )$ according to experimental procedure~\cite{Aaij:2016kfs}. If we exclude $J/\psi$ and $\psi(2S)$ contributions following the conditions $\sqrt{|M^2_1x_{ij} - m^2(Res)|} < 100\,\textrm{MeV}$ and the $\phi(1020)$ contribution with the condition $\sqrt{|M^2_1x_{ij} - m^2(\phi(1020))|} < 70\,\textrm{MeV}$ according to the ~\cite{4mu2021}, we get the second value for the $\textrm{Br}\,\left (\bar B_s \to \mu^+ \mu^- e^+ e^- \right )$ decay:
$$
{\textrm{Br}\,\left (\bar B_s \to \mu^+ \mu^- e^+ e^- \right ) \approx (2.8\pm 0.5)*10^{-10}}.
$$
Based on the our numerical predictions of the $\textrm{Br}\,\left (\bar B_s \to \mu^+ \mu^- e^+ e^- \right )$ it is possible to use the following
estimate for the branching ratio of $\bar B_s \to \mu^+ \mu^- \mu^+ \mu^-$ decay.
If we exclude the $\phi(1020)$ resonances contribution, in accordance with \cite{Dincer:2003zq}  we get
$$\textrm{Br}\,\left (\bar B_s \to \mu^+ \mu^- e^+ e^- \right ) : \textrm{Br}\,\left (\bar B_s \to \mu^+ \mu^- \mu^+ \mu^- \right ) = 3 : 1. $$
We have 
$$ 
{\textrm{Br}\,\left (\bar B_s \to \mu^+ \mu^- \mu^+ \mu^- \right ) \sim 10^{-10}}.
$$
This estimation does not contradict to the experimental upper limit \cite{4mu2021}:
 $${\textrm{Br}_{Exp}\,\left (\bar B_s \to \mu^+ \mu^- \mu^+ \mu^- \right )\,\le\, 8.6\,* 10^{-10}}$$
 and is consistent with the estimation from \cite{Danilina:2018uzr}.

\section{Differential distributions}

We study the set of differential characteristics for the $\bar B_s \to \mu^+ \mu^- e^+ e^-$ decay, that demonstrate the features of this rare four – leptonic decay. We take into account contributions of $\phi(1020)$, $\psi(3770)$, $\psi(4040)$, $\psi(4160)$, $\psi(4415)$, $\rho^0(770)$ and $\omega(782)$ resonances and “tails” from the   $ {J/\psi}$ and ${\psi(2S)}$ resonances. 

\begin{figure}[h!]
\begin{minipage}[h]{0.47\linewidth}
\center{\includegraphics[width=1\linewidth]{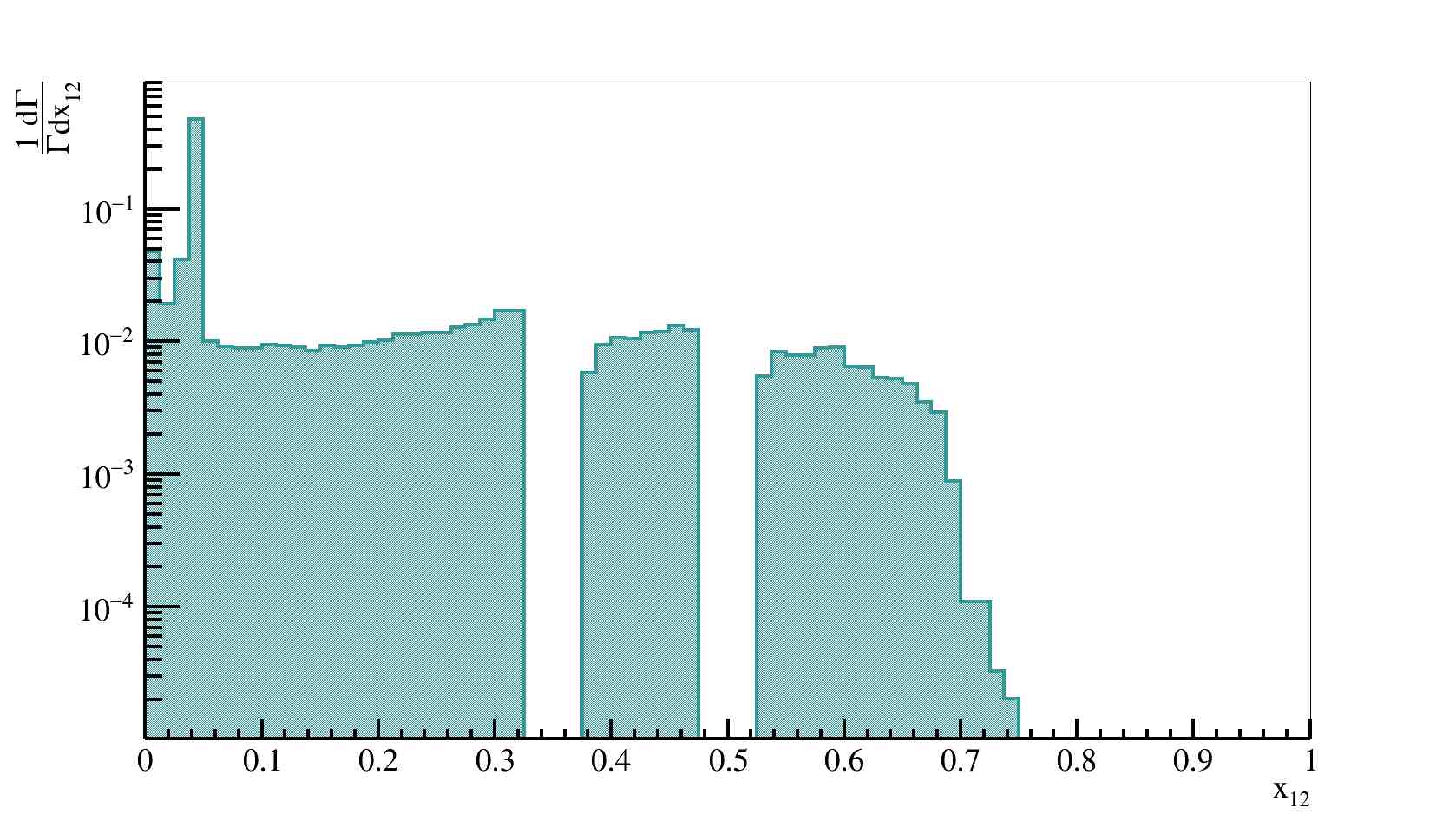}} 1) \\
\end{minipage}
\hfill
\begin{minipage}[h]{0.47\linewidth}
\center{\includegraphics[width=1\linewidth]{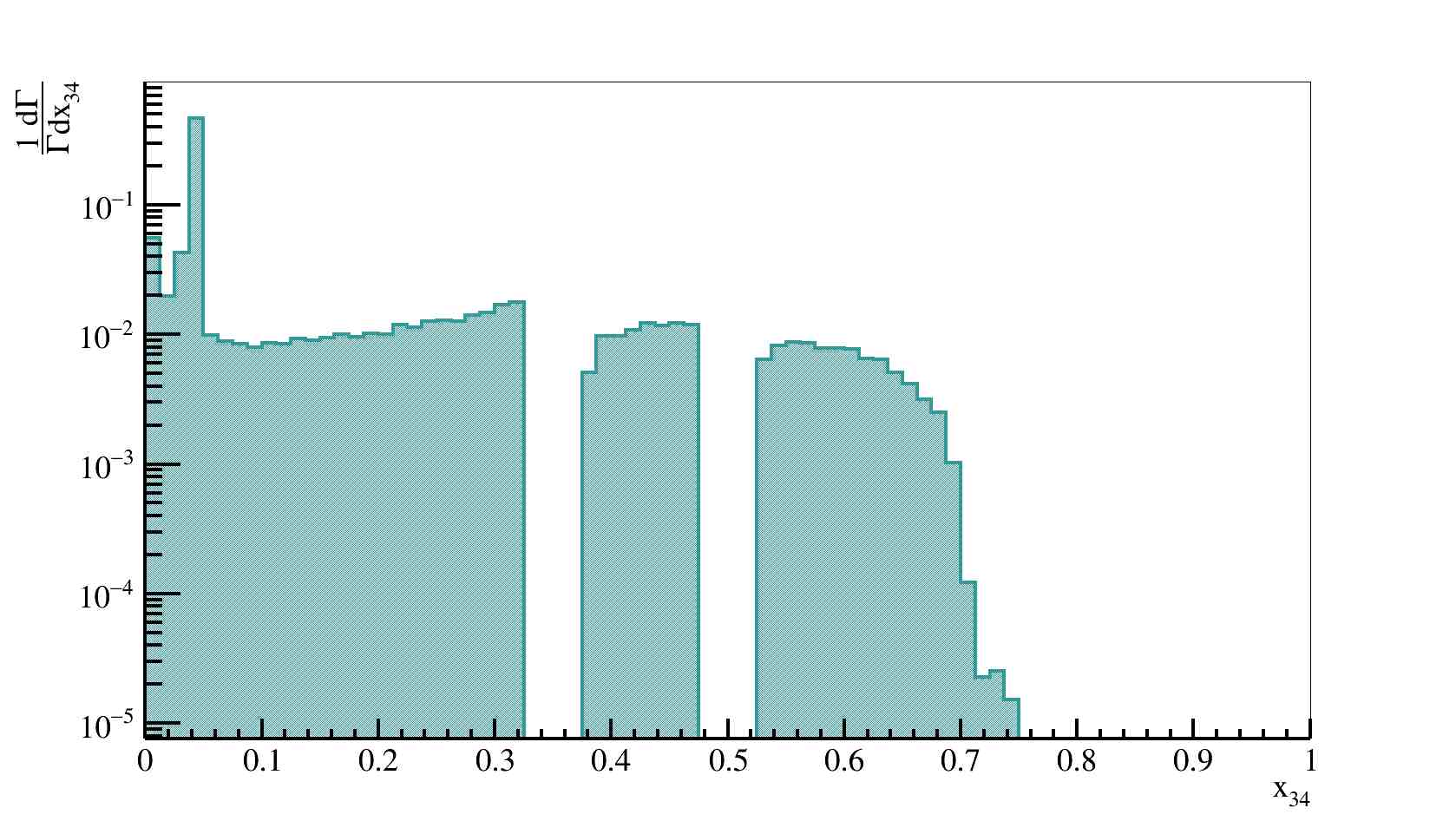}} 2)\\
\end{minipage}
\vfill
\begin{minipage}[h]{0.47\linewidth}
\center{\includegraphics[width=1\linewidth]{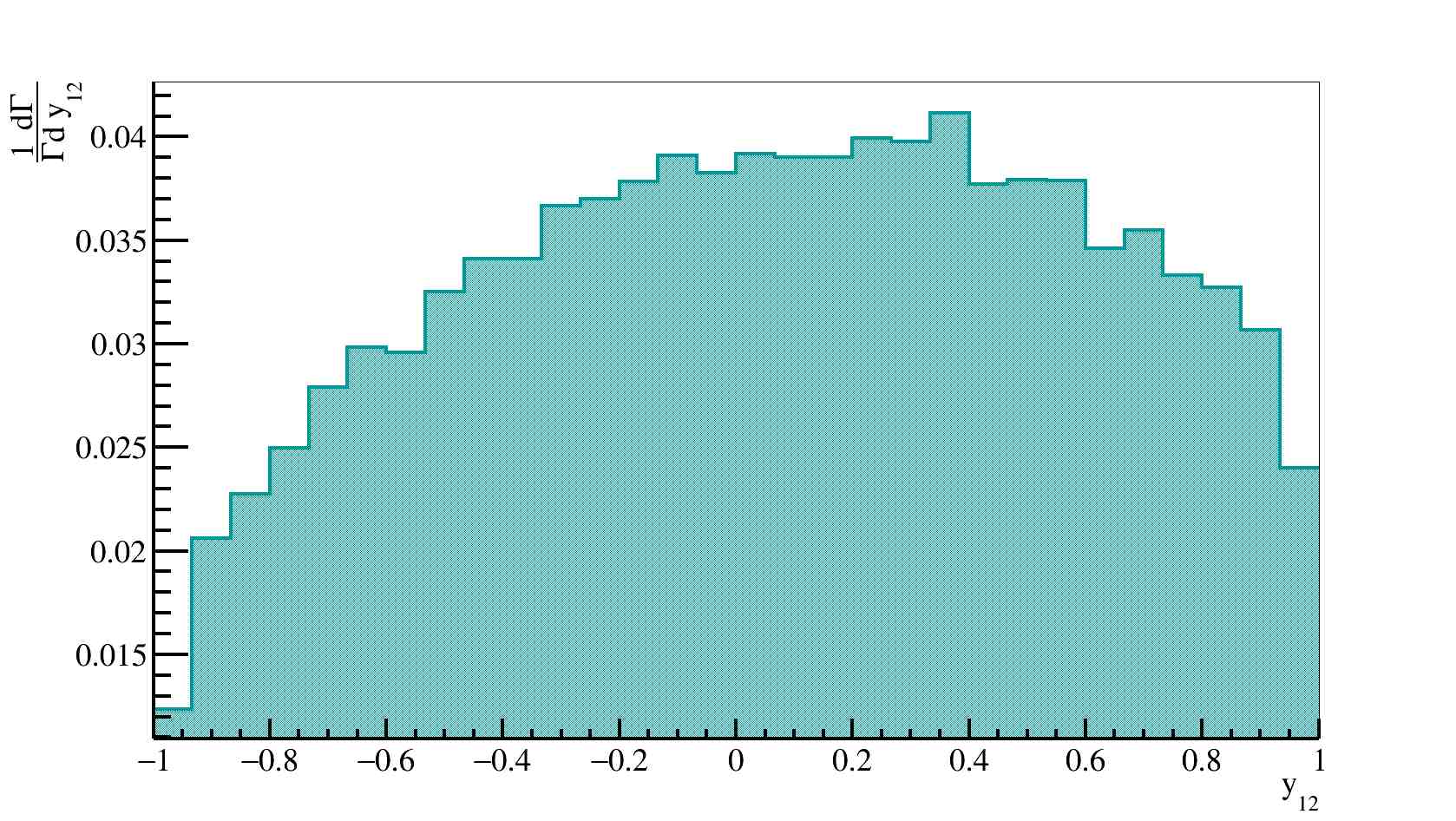}} 3) \\
\end{minipage}
\hfill
\begin{minipage}[h]{0.47\linewidth}
\center{\includegraphics[width=1\linewidth]{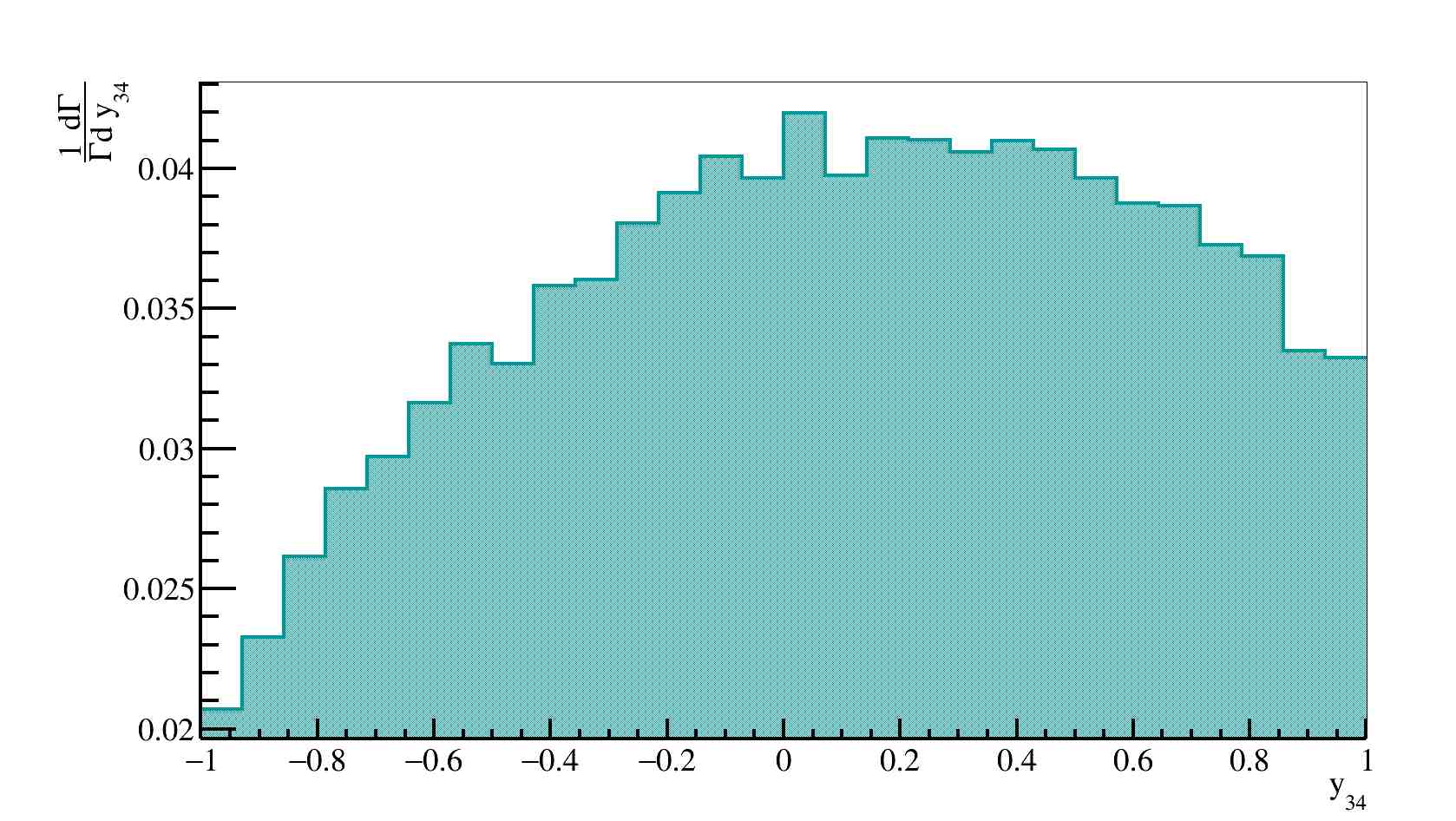}} 4)\\
\end{minipage}
\vfill
\caption {1) $x_{12}$  - distribution for the $\mu^+\mu^-$ - pair; 2) $x_{34}$  - distribution for the $e^+e^-$ - pair ; 3) $cos(\theta_{12})$ - distribution; 4) $cos(\theta_{34})$ - distribution.}
\label {ris:image}
\end{figure}

We consider differential distributions for the decay $\bar B_s \to \mu^+ \mu^- e^+ e^-$.
One-dimensional differential distribution by $x_{12}$
and $x_{34}$ are given in
Fig.~\ref{ris:image}(1) and  Fig.~\ref{ris:image}(2)
respectively. Here we exclude ${J/\psi}$ resonance region in accordance with conditions $\sqrt{|M^2_1x_{12} - m^2(J/\psi)|} < 100\,\textrm{MeV}$. We have the same kinematical cut for the $\psi(2S)$ meson: $\sqrt{|M^2_1x_{12} - m^2(\psi(2S))|} < 100\,\textrm{MeV}$. These restrictions selected according to experimental cuts~\cite{4mu2021}. The same cuts we use for the $x_{34}$.
For the differential distributions the $\phi(1020)$ resonance is not excluded.
In the Fig.~\ref{ris:image}(1) it is
shown a photon pole for $x_{12} \to x_{12\, min}= (2 m_\mu /M_1)^2 =
0.0016$ and a peak from the $\phi(1020)$ resonance for $x_{12} \to
(M_\phi  /M_1)^2 \approx 0.037$. Due to the fact that the  $\phi(1020)$ meson contribution is very high, the contributions from other resonances near $1\, \textrm{GeV}$ ($\rho^0(770)$ and $\omega(782)$ mesons) are not visible in Fig.~\ref{ris:image}(1). The distributions
by $x_{34}$ in Fig.~\ref{ris:image}(2) also reflect  $\phi(1020)$ resonance contribution. It is generally symmetrical to the Fig.~\ref{ris:image}(1). 

The distributions of the angular variables $y_{12} = \cos\theta_{12}$ and $y_{34} = \cos\theta_{34}$ are presented in Fig.~\ref{ris:image}(3)
and Fig.~\ref{ris:image}(2) respectively. Here $\theta_{12}$ is the angle between the
propagation directions of the $\mu^+$ and $B_s$ in the rest frame of the 
$\mu^+\mu^-$ -- pair. The angel $\theta_{34}$ is the angle between
the propagation directions of $e^{\,+}$ and $B_s$ in the rest
frame of the $e^+e^-$ pair.

\begin{figure}[h!]
\begin{minipage}[h]{1.0\linewidth}
\center{\includegraphics[width=1.0\linewidth]{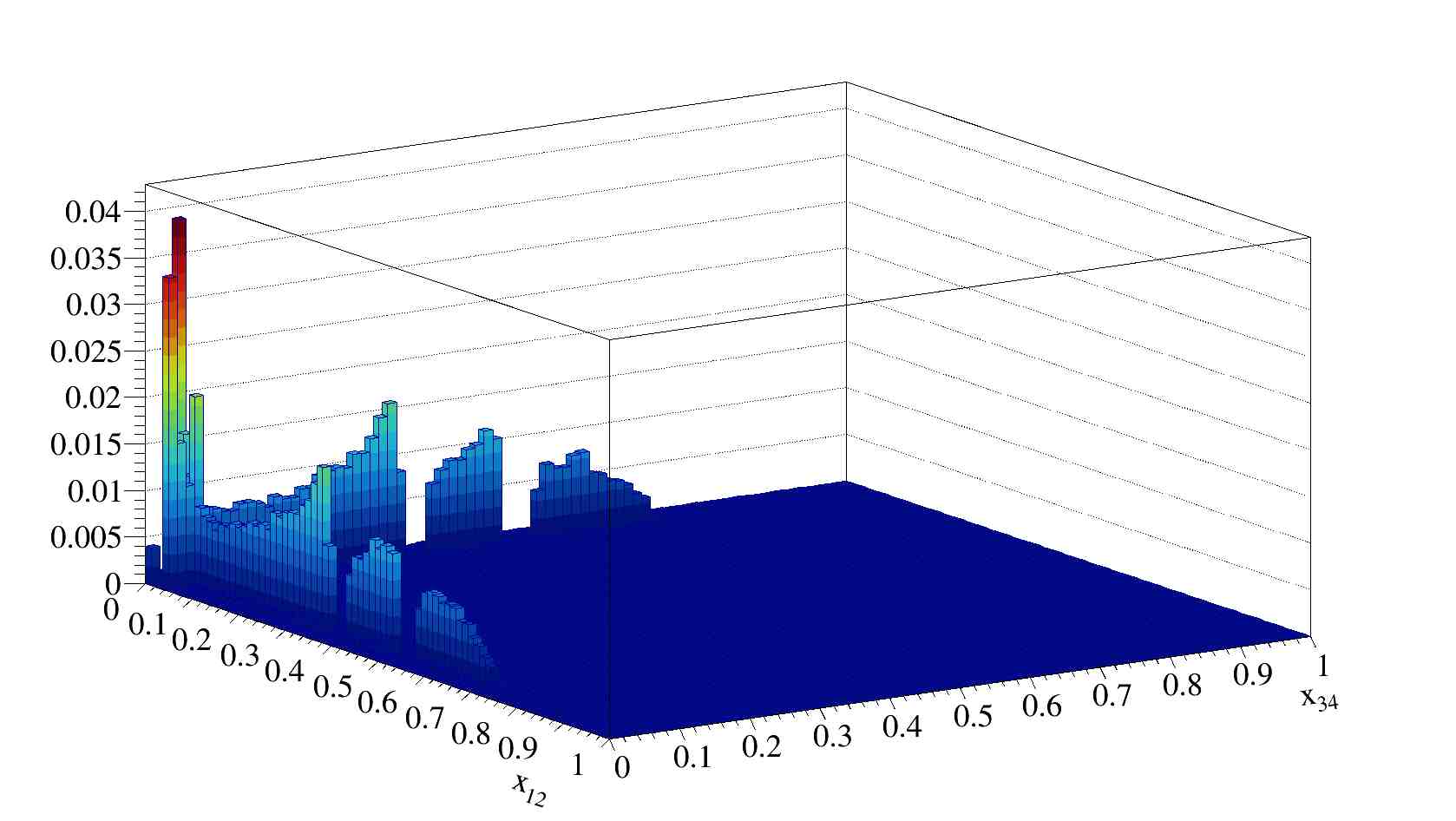}} \\
\end{minipage}
\caption{Double differential
  distribution $\displaystyle \frac{d^2\textrm{Br}
    \left (\bar B_s \to \mu^+ \mu^- e^+ e^- \right
    )}{dx_{12}\, d x_{34}}$}
    \label{ris:q2k2}
\end{figure}

The
importance of the pole contribution becomes obvious when we analyze the double differential distribution $\displaystyle \frac{d^2\textrm{Br}
    \left (\bar B_s \to \mu^+ \mu^- e^+ e^- \right
    )}{dx_{12}\, d x_{34}}$, which is presented in Fig.~\ref{ris:q2k2}.
    The Fig.~\ref{ris:q2k2} features the $\phi(1020)$ resonance
in the $x_{12}$
and $x_{34}$ channels and the “tails” from $ {J/\psi}$ and ${\psi(2S)}$  resonances contributions.

The forward -- fackward lepton asymmetries are very
sensitive to BSM physics. The asymmetry
shape provides the fundamental information about the behaviour of the $B_s$ amplitude and effects of intermediate resonances. 
For the decay $\bar B_s \to \mu^+ \mu^- e^+ e^-$ it is possible to forward -- fackward lepton
asymmetries $A^{(\bar{B_s})}_{FB}(x_{12})$ and $A^{(\bar{B_s})}_{FB}(x_{34})$
according to

$$
\label{Afbq2-def}
A^{(\bar{B_s})}_{FB}(x_{12})\, =\,
\frac{\int\limits_0^1
d y_{12}\,\frac{\displaystyle d^2\,\Gamma \left (\bar B_s \to \mu^+ \mu^- e^+ e^- \right )}{\displaystyle d x_{12}\, d y_{12}}\,-\,
\int\limits_{-1}^0
d y_{12}\,\frac{\displaystyle d^2\,\Gamma \left (\bar B_s \to \mu^+ \mu^- e^+ e^- \right )}{\displaystyle d x_{12}\, d y_{12}}
}
{ \frac{\displaystyle d\,\Gamma \left (\bar B_s \to \mu^+ \mu^- e^+ e^- \right )}{\displaystyle d x_{12}}}
$$
and
$$
\label{Afbk2-def}
A^{(\bar{B_s})}_{FB}(x_{34})\, =\,
\frac{\int\limits_0^1
d y_{34}\,\frac{\displaystyle d^2\,\Gamma \left (\bar B_s \to \mu^+ \mu^- e^+ e^- \right )}{\displaystyle d x_{34}\, d y_{34}}\,-\,
\int\limits_{-1}^0
d y_{34}\,\frac{\displaystyle d^2\,\Gamma \left (\bar B_s \to \mu^+ \mu^- e^+ e^- \right )}{\displaystyle d x_{34}\, d y_{34}}
}
{ \frac{\displaystyle d\,\Gamma \left (\bar B_s \to \mu^+ \mu^- e^+ e^- \right )}{\displaystyle d x_{34}}}.
$$

\begin{figure}[h!]
\begin{minipage}[h]{0.47\linewidth}
\center{\includegraphics[width=1.2\linewidth]{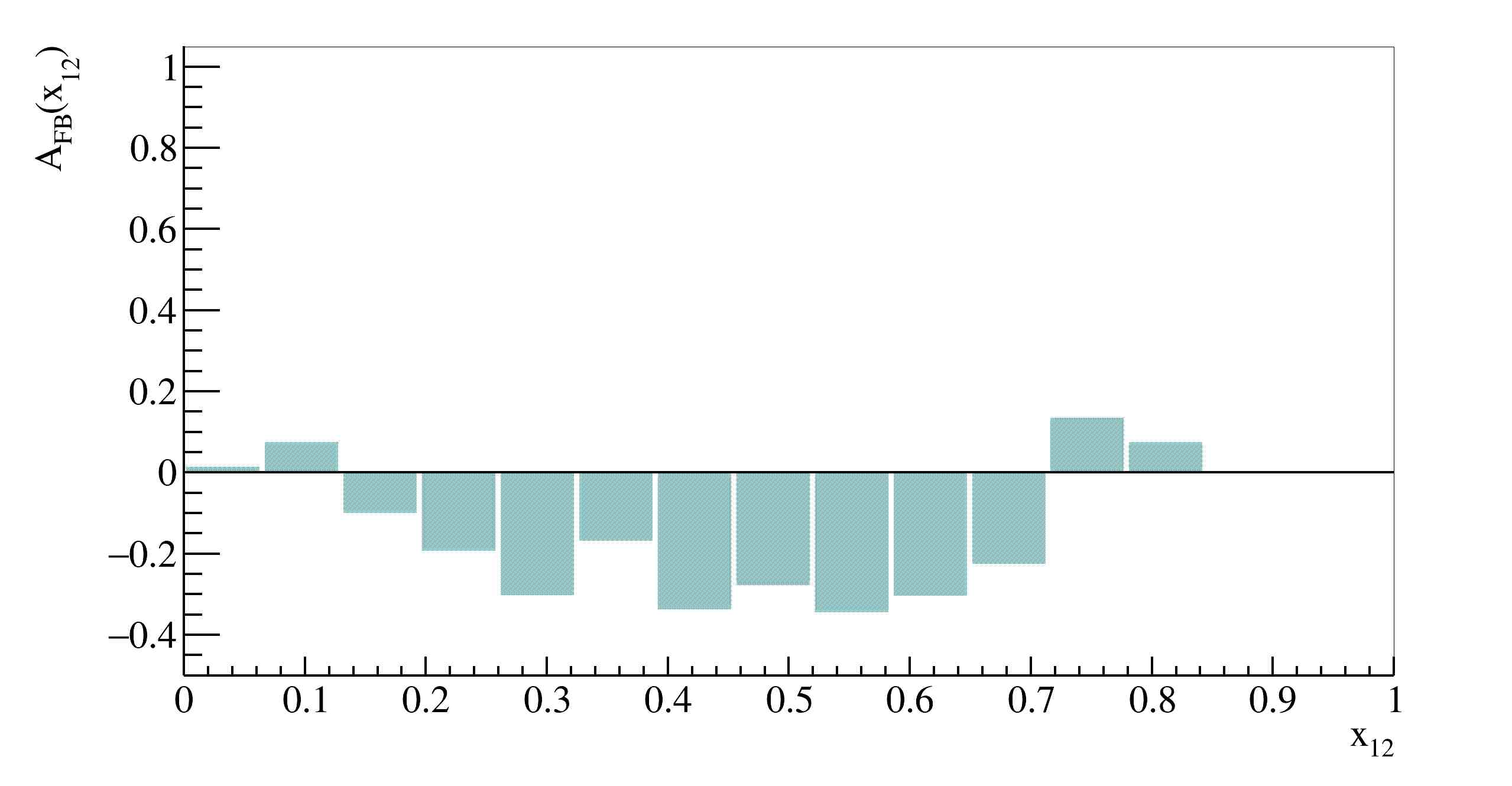}} 1) \\
\end{minipage}
\hfill
\begin{minipage}[h]{0.47\linewidth}
\center{\includegraphics[width=1.2\linewidth]{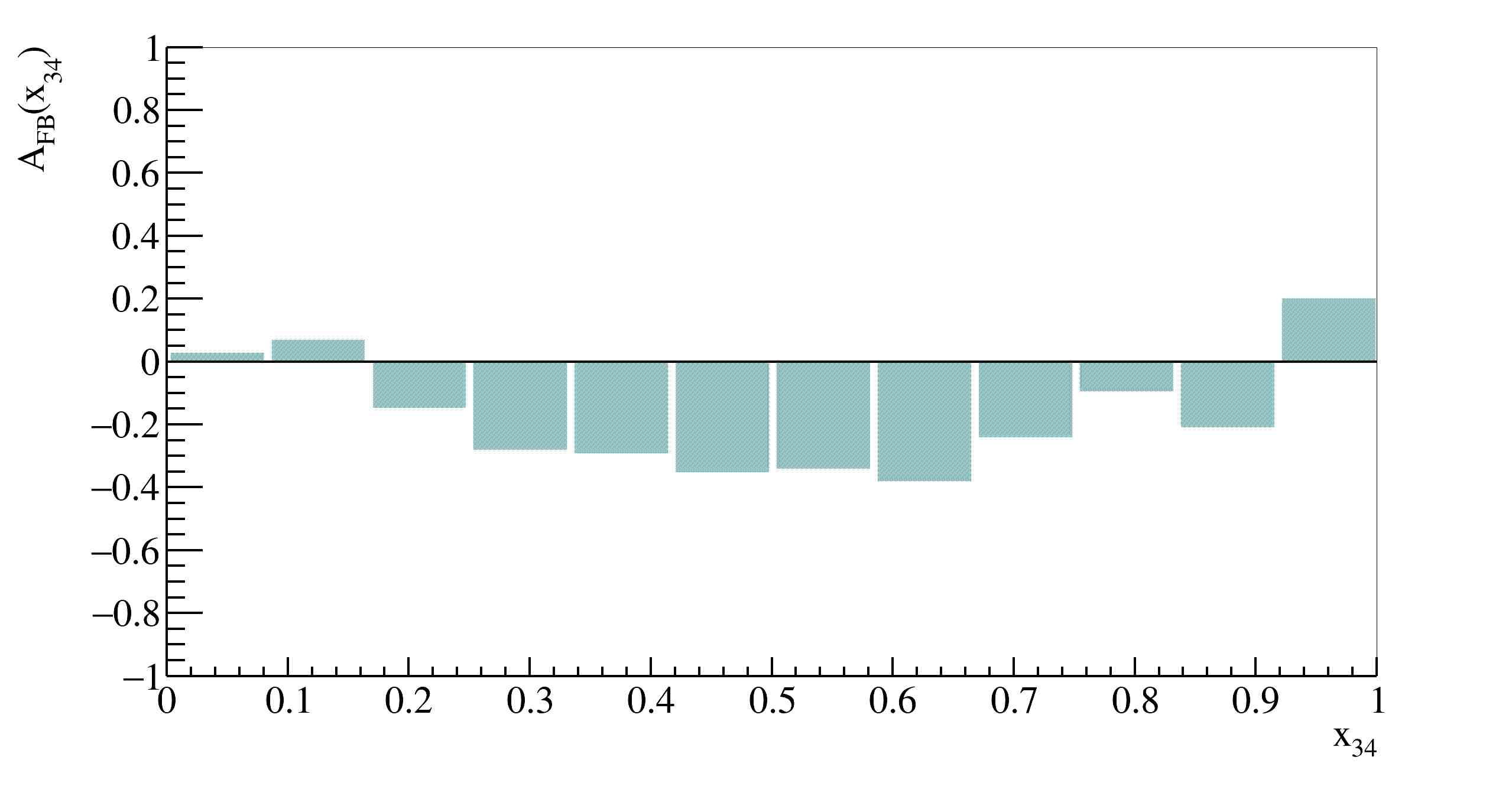}} 2)\\
\end{minipage}
\vfill
\caption{1)The forward–backward leptonic asymmetry for the $\mu^+\mu^-$ - pair; \\ 2) The forward–backward leptonic asymmetry for the $e^+e^-$ - pair.}
\label {ris:FBA}
\end{figure}

These
asymmetries are shown in Fig.~\ref{ris:FBA}. The shape of asymmetries makes good sense in both channels and very similar to the shape of the asymmetries for the rare semileptonic decays of $B_{d,s}$ -- mesons. It passes through zero in the region of small $x_{12}$ (or $x_{34}$) and  reflects the influence of the $\phi(1020)$. In the region of big $x_{12}$ (or $x_{34}$) this shape demonstrates the relative signs between  the higher exited charmonium states.

\section{Conclusion}
\begin{itemize}
 \item In the framework of the Standard Model we present the predictions for the $\textrm{Br}\,\left (\bar B_s \to \mu^+ \mu^- e^+ e^- \right )$. We take into account resonant contributions $\phi(1020)$, $\psi(3770)$, $\psi(4040)$, $\psi(4160)$, $\psi(4415)$, $\rho(770)$ and $\omega(782)$. Also we consider  “tails” contributions from $J/\psi$ and $\psi(2S)$ resonances, non -- resonant contribution from $b\,\bar{b}$ - pairs, the weak annihilation contribution and the bremsstrahlung. For the excluded $J/\psi$ and $\psi(2S)$ resonances in accorde with ~\cite{Aaij:2016kfs,Aaij:2013lla} we have
$$
{\textrm{Br}\,\left (\bar B_s \to \mu^+ \mu^- e^+ e^- \right ) \approx (61\pm 12)*10^{-10}};
$$
If we exclude the contributions of $\phi(1020)$, $J/\psi$ and $\psi(2S)$ in accordance with \cite{4mu2021}, we obtain 
$$
{\textrm{Br}\,\left (\bar B_s \to \mu^+ \mu^- e^+ e^- \right ) \approx (2.5\pm 0.5)*10^{-10}};
$$
\item  We provide the estimation for the branching ratio of ${\bar B_s \to \mu^+ \mu^- \mu^+ \mu^- }$ decay at the level
$$ 
{\textrm{Br}\,\left (\bar B_s \to \mu^+ \mu^- \mu^+ \mu^- \right ) \sim 10^{-10}}.
$$
This prediction is consistent with the recent experimental upper limit from \cite{4mu2021};
\item Using the EvtGen-based 
generator model, we obtain the set of differential
distributions for the $\bar B_s \to \mu^+ \mu^- e^+ e^-$ decay.

\end{itemize}


\section*{Acknowledgements}
This paper is dedicated to the memory of Konstantin Toms, dear colleague and friend,  who took interest in this work
and did not live to see it published.\\
The authors would like to thank I.~M.~Belyaev (ITEP),  E.~E.~Boos (SINP
MSU), L.~V.~Dudko (SINP MSU), D.~I.~Melikhov (SINP MSU),  V.~Yu.~Yegorychev (ITEP), and  D.~V.~Savrina (ITEP, SINP MSU) for
fruitful discussions which improved the current work significantly.

N. Nikitin was supported by RFBR under project 19-52-15022.

A.~Danilina is grateful to the Basis Foundation for her stipend for
Ph.D. students and express her gratitude to the Olga Igonkina Foundation for supporting this work.

\appendix
\section{Kinematics of four-lepton decays}
\label{sec;kinemat4}

Denote the four-momenta of  the final leptons in four-leptonic decays
of $B$--mesons as $k_i$, $i =\{1,\, 2,\, 3,\, 4\}$.  Let 
$$
q = k_1 + k_2; \quad
k = k_3 + k_4; \quad
\tilde q = k_1 + k_4; \quad
\tilde k = k_2 + k_3; \quad
p = k_1 + k_2 + k_3 + k_4,
$$ 
where $p$ is the four-momentum of the $B$--meson and $p^2 = M^2_1$. For the
calculations below it is suitable to use the dimensionless
variables: 
$$
x_{12} = \frac{q^2}{M_1^2}, \quad
x_{34} = \frac{k^2}{M_1^2}, \quad
x_{14} = \frac{{\tilde q}^2}{M_1^2}, \quad
x_{23} = \frac{{\tilde k}^2}{M_1^2}.
$$
By common notation, $x_{ij} = (k_i + k_j)^2/M^2_1$. Hence $x_{ij} =
x_{ji}$. The leptons may be considered as massless in almost all of
the calculations of the present work, i.e., $k_i^2=0$. However during
the calculation of the bremsstrahlung contribution in the area $q^2$ and $k^2$
, it is necessary to take into account
the dependence of the bremsstrahlung matrix element and phase space on values of $m_e$ and $m_\mu$.

Let us find the intervals for $x_{ij}$ using the inequality
$(p_1 p_2) \ge \sqrt{p_1^2\, p_2^2}$; then any $x_{ij} \ge (\hat{m_i} + \hat{m_j})^2$. On the
other hand,
\begin{eqnarray}
1 & =&\frac{p^2}{M_1^2}\, =\,\frac{(q+k)^2}{M^2_1}\,\ge \,\frac{(\sqrt{q^2} + \sqrt{k^2})^2}{M_1^2}\, =\,\Big (\sqrt{x_{12}} + \sqrt{x_{34}} \Big)^2.
\nonumber
\end{eqnarray}
As $(\hat{m_3} + \hat{m_4})^2 \le x_{34}$, then  $x_{12} \le 1$, so $x_{12} \in [0,\, 1]$. The
upper limit of the variable $x_{34}$ depends on the value of $x_{12}$:
$$
x_{34}\, =\,\frac{(p - q)^2}{M_1^2}\,\le\,\frac{(M_1 - \sqrt{q^2})^2}{M_1^2}\, =\,\left ( 1 - \sqrt{x_{12}}\right )^2.
$$
Thus for a fixed value of $x_{12}$ the variable $x_{34} \in \left [ (\hat{m_3} + \hat{m_4})^2,\,\left ( 1 - \sqrt{x_{12}}\right )^2 \right ]$.

\begin{figure}[h!]
\begin{center}
\includegraphics[width=11.0cm]{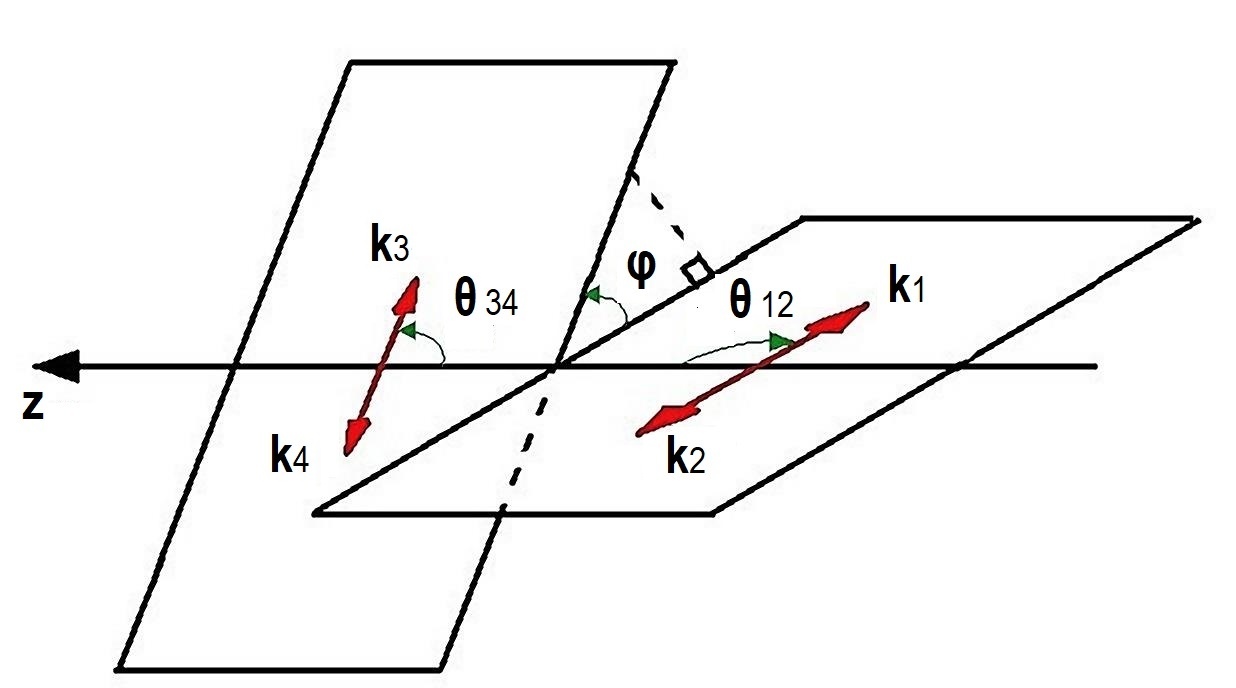} 
\end{center}
\caption{\protect\label{fig:cinematic} 
Kinematics of the decay $\bar B_s(p) \,\to\, \mu^+(k_1)\,\mu^-(k_2)\,  e^+(k_3)\,e^-(k_4)$.  Angle $\theta_{12}$ is
defined in the rest frame of $\mu^+\,\mu^-$--pair;  angle
$\theta_{34}$ is defined in the rest frame of $e^+\,e^-$--pair; angle $\varphi$ is defined
in the rest frame of $\bar{B_s}$--meson.
}
\end{figure}

Consider the kinematics of the decay
$ 
\bar B_s(p) \,\to\, \mu^+(k_1)\,\mu^-(k_2)\,  e^+(k_3)\,e^-(k_4) 
$. We define an angle
$\theta_{12}$ between the momentum of $\mu^+$ and
the direction of the $\bar{B_s}$--meson ($z$--axis) in the rest frame of the
$\mu^+\mu^-$ pair, and another angle $\theta_{34}$ between the
direction of the $e^+$ and the direction of the $\bar{B_s}$ -- meson
($z$--axis) in the rest frame of $e^+ e^-$ -- pair. Then
\begin{eqnarray}
\label{costheta-equation}
y_{12}\,\equiv\,\cos\theta_{12} &=& \frac{1}{\lambda^{1/2} (1,\, x_{12},\, x_{34})}\,\left ( x_{23} + x_{24} - x_{13} - x_{14}\right ),
\\
y_{34}\,\equiv\,\cos\theta_{34} &=& \frac{1}{\lambda^{1/2} (1,\, x_{12},\, x_{34})}\,\left ( x_{14} + x_{24} - x_{13} - x_{23}\right ),
\nonumber
\end{eqnarray}
where $\lambda (a, \, b,\, c) = a^2 + b^2 + c^2 - 2ab - 2ac - 2bc$, the 
triangle function. Angles $\theta_{12} \in \left [0,\,\pi \right ]$
and $\theta_{34} \in \left [0,\,\pi \right ]$. Hence $y_{12} \in
[-1,\, 1]$ and $y_{34} \in [-1,\, 1]$. Angles are measured relative to
$z$--axis. Also let us define an angle $\varphi \in [0,\, 2 \pi)$ in
the rest frame of the $\bar{B_s}$--meson between the planes which are set by the pairs
of vectors $({\bf k}_1,\, {\bf k}_2)$ and $({\bf k}_3,\, {\bf
  k}_4)$.  Introduce a vector ${\bf  a}_1 = {\bf k}_1\,\times\,
{\bf k}_2$, perpendicular to the plane $({\bf k}_1,\, {\bf k}_2)$, and 
vector ${\bf  a}_3 = {\bf k}_{\, 4}\,\times\, {\bf k}_{\, 3}$, which are
normal to the plane $({\bf k}_3,\, {\bf k}_4)$. Then
$$
\cos\varphi\, =\,\frac{\Big ( {\bf  a}_1,\, {\bf  a}_3\Big )}{|{\bf  a}_1|\, |{\bf  a}_3|}.
$$

It is suitable to choose $x_{12}$, $x_{34}$, $y_{12}$, $y_{34}$, and
$\varphi$ as independent integration variables. Then the four body phase space has the form
 
\begin{eqnarray}
\label{dPhi1234}
\fl
 \,\,\,\,\,\,\,\,\,d\Phi_4=                                                    \frac{M_1^4}{2^{14}\, \pi^6}\,
\lambda^{1/2} \left ( 1,\,  x_{12},\, x_{34}\right )\,\sqrt{1\, -\,\frac{4 {\hat m}_\mu^2}{x_{12}}}\,\, \sqrt{1\, -\,\frac{4{\hat m}_{e}^2}{x_{34}}}
d x_{12}\, d x_{34}\, d y_{12}\, d y_{34}\, d \varphi, 
\end{eqnarray} 
where ${\hat m}_\mu = m_\mu /M_1$ and  ${\hat m}_{e} = m_{e}/M_1$.

This paper use notations almost  identical
to the notations of Ref. \cite{Barker:2002ib}, except for 
$y_{ij}$, which here have the opposite sign compared to
Ref. \cite{Barker:2002ib}.

\section{Dimensionless non -- zero 
functions $a^{(ij)}, b^{(ij)}$, $c^{(ij)}$, $d^{(ij)}$, $f^{(ij)}$ and $g^{(ij)}$}
Here we define the dimensionless functions from the (\ref{Amp}) decay amplitude.
\label{sec;ABC}
\begin{eqnarray}
\label{abc_x12x34}
\fl a^{(VV)}(x_{12}\, x_{34}) = \frac{1}{M_1^2}\,\Bigg[ \frac{4\,\hat m_{b}\, C_{7\gamma}(\mu)}{x_{12},\, x_{34}}\,\Bigg(\frac{1}{2}(F_{TV}(q^2,k^2)+F_{TV}(k^2,q^2)) -\nonumber\\ - \frac{\hat M_{2}\,\hat f^{em}_\phi}{x_{34} - \hat M_{2}^2+i\hat \Gamma_{2} \hat M_{2}}\,T_1(q^2) - \frac{\hat M_{2}\,\hat f^{em}_\phi}{x_{12} - \hat M_{2}^2+i\hat \Gamma_{2} \hat M_{2}}\,T_1(k^2)\Bigg) +\nonumber\\ +\,
\frac{C_{9V}(q^2,\,\mu)}{x_{34}}\,\Bigg(F_{V}(q^2,k^2)  - \frac{2\hat M_{2}}{1 + \hat M_{2}}\,\frac{V(q^2)\,\hat f^{em}_\phi}{x_{34} - \hat M_{2}^2+i\hat \Gamma_{2} \hat M_{2}}\,\Bigg) +\nonumber\\ +\,
\frac{C_{9V}(k^2,\,\mu)}{x_{12}}\,\Bigg(F_{V}(k^2,q^2)  - \frac{2\hat M_{2}}{1 + \hat M_{2}}\,\frac{V(k^2)\,\hat f^{em}_\phi}{x_{12} - \hat M_{2}^2+i\hat \Gamma_{2} \hat M_{2}}\,\Bigg)\Bigg]\,;
\nonumber
\end{eqnarray}
\\
\begin{eqnarray}
\label{abc_x12x34}
\fl a^{(VA)}(x_{12},\, x_{34}) = \frac{1}{M_1^2}\,\frac{C_{10A}}{x_{12}}\,\Bigg[ \,F_{V}(k^2,q^2)  - \frac{2\hat M_{2}}{1 + \hat M_{2}}\,\frac{V(k^2)\,\hat f^{em}_\phi}{x_{12} - \hat M_{2}^2+i\hat \Gamma_{2} \hat M_{2}}\,\Bigg]\,;
\nonumber
\end{eqnarray}
\\
\begin{eqnarray}
\label{abc_x12x34}
\fl a^{(AV)}(x_{12},\, x_{34}) = \frac{1}{M_1^2}\,\frac{C_{10A}}{x_{34}}\,\Bigg[ \,F_{V}(q^2,k^2)  - \frac{2\hat M_{2}}{1 + \hat M_{2}}\,\frac{V(q^2)\,\hat f^{em}_\phi}{x_{34} - \hat M_{2}^2+i\hat \Gamma_{2} \hat M_{2}}\,\Bigg]\,;
\nonumber
\end{eqnarray}
\\
\begin{eqnarray}
\fl b^{(VV)}(x_{12},\, x_{34}) = \frac{1}{M_1^2}\,\Bigg[ \frac{2\,\hat m_{b}\, C_{7\gamma}(\mu)}{x_{12},\, x_{34}}\,\Bigg(\frac{1 - x_{12} -x_{34}}{2}(F_{TA}(q^2,k^2)+F_{TA}(k^2,q^2)) -\nonumber\\ - \frac{\hat M_{2}\,(1 - \hat M_{2}^2)\,\hat f^{em}_\phi}{x_{34} - \hat M_{2}^2+i\hat \Gamma_{2} \hat M_{2}}\,T_2(q^2) - \frac{\hat M_{2}\,(1 - \hat M_{2}^2)\,\hat f^{em}_\phi}{x_{12} - \hat M_{2}^2+i\hat \Gamma_{2} \hat M_{2}}\,T_2(k^2)\Bigg) +\nonumber\\ +\,
\frac{C_{9V}(q^2,\,\mu)}{x_{34}}\,\Bigg(\frac{1}{2}(1 - x_{12} -x_{34})F_{A}(q^2,k^2)  - \frac{\hat M_{2}\,(1 + \hat M_{2})\,\hat f^{em}_\phi}{x_{34} - \hat M_{2}^2+i\hat \Gamma_{2} \hat M_{2}}\,A_1(q^2)\Bigg) +\nonumber\\ +\,
\frac{C_{9V}(k^2,\,\mu)}{x_{12}}\,\Bigg(\frac{1}{2}(1 - x_{12} -x_{34})F_{A}(k^2,q^2)  - \frac{\hat M_{2}\,(1 + \hat M_{2})\,\hat f^{em}_\phi}{x_{12} - \hat M_{2}^2+i\hat \Gamma_{2} \hat M_{2}}\,A_1(k^2)\Bigg)\Bigg]\,;
\nonumber
\end{eqnarray}
\\
\begin{eqnarray}
\label{abc_x12x34}
\fl b^{(VA)}(x_{12},\, x_{34}) = \frac{1}{M_1^2}\,\frac{C_{10A}(\mu)}{x_{12}}\,\Bigg[ \,\frac{1\,-x_{12}-x_{34}}{2}\,F_{A}(k^2,q^2)  - \,\frac{\hat M_{2}\,(1 + \hat M_{2})\,\hat f^{em}_\phi}{x_{12} - \hat M_{2}^2+i\hat \Gamma_{2} \hat M_{2}}\,A_1(k^2)\,\Bigg]\,;
\nonumber
\end{eqnarray}
\\
\begin{eqnarray}
\label{abc_x12x34}
\fl b^{(AV)}(x_{12},\, x_{34}) = \frac{1}{M_1^2}\,\frac{C_{10A}(\mu)}{x_{34}}\,\Bigg[ \,\frac{1\,-x_{12}-x_{34}}{2}\,F_{A}(q^2,k^2)  - \,\frac{\hat M_{2}\,(1 + \hat M_{2})\,\hat f^{em}_\phi}{x_{34} - \hat M_{2}^2+i\hat \Gamma_{2} \hat M_{2}}\,A_1(q^2)\,\Bigg]\,;
\nonumber
\end{eqnarray}
\\
\begin{eqnarray}
\fl c^{(VV)}(x_{12},\, x_{34}) = \frac{1}{M_1^2}\,\Bigg[ \frac{2\,\hat m_{b}\, C_{7\gamma}(\mu)}{x_{12},\, x_{34}}\,\Bigg(\frac{1}{2}(F_{TA}(q^2,k^2)+F_{TA}(k^2,q^2)) - \nonumber\\ - \frac{\hat M_{2}\,\hat f^{em}_\phi}{x_{34} -  \hat M_{2}^2+i\hat \Gamma_{2} \hat M_{2}}\,\Bigg(T_2(q^2) + \frac{T_3(q^2)\,x_{12}}{(1 - \hat M_{2}^2)\,}\Bigg) -\nonumber\\ - \frac{\hat M_{2}\,\hat f^{em}_\phi}{x_{12} - \hat M_{2}^2+i\hat \Gamma_{2} \hat M_{2}}\,\Bigg(T_2(k^2) + \frac{T_3(k^2)\,x_{34}}{(1 - \hat M_{2}^2)\,}\Bigg)\Bigg) +\nonumber\\ +\,
\frac{C_{9V}(q^2,\,\mu)}{x_{34}}\,\Bigg(\frac{1}{2}F_{A}(q^2,k^2)  - \frac{\hat M_{2}}{(1 + \hat M_{2})}\frac{A_2(q^2)\hat f^{em}_\phi}{x_{34} - \hat M_{2}^2+i\hat \Gamma_{2} \hat M_{2}}\Bigg) +\nonumber\\ +\,
\frac{C_{9V}(k^2,\,\mu)}{x_{12}}\,\Bigg(\frac{1}{2}F_{A}(k^2,q^2)  - \frac{\hat M_{2}}{(1 + \hat M_{2})}\frac{A_2(k^2)\hat f^{em}_\phi}{x_{12} - \hat M_{2}^2+i\hat \Gamma_{2} \hat M_{2}}\Bigg)\Bigg]\,;
\nonumber
\end{eqnarray}
\\
\begin{eqnarray}
\label{abc_x12x34}
\fl c^{(VA)}(x_{12},\, x_{34}) = \frac{1}{M_1^2}\,\frac{C_{10A}(\mu)}{x_{12}}\,\Bigg[ \,\frac{1}{2}\,F_{A}(k^2,q^2)  - \frac{\hat M_{2}}{1 + \hat M_{2}}\,\frac{A_2(k^2)\,\hat f^{em}_\phi}{x_{12} - \hat M_{2}^2+i\hat \Gamma_{2} \hat M_{2}}\,\Bigg]\,;
\nonumber
\end{eqnarray}
\\
\begin{eqnarray}
\label{abc_x12x34}
\fl c^{(AV)}(x_{12},\, x_{34}) = \frac{1}{M_1^2}\,\frac{C_{10A}(\mu)}{x_{34}}\,\Bigg[ \,\frac{1}{2}\,F_{A}(q^2,k^2)  - \frac{\hat M_{2}}{1 + \hat M_{2}}\,\frac{A_2(q^2)\,\hat f^{em}_\phi}{x_{34} - \hat M_{2}^2+i\hat \Gamma_{2} \hat M_{2}}\,\Bigg]\,;
\nonumber
\end{eqnarray}
\\
\begin{eqnarray}
\label{abc_x12x34}
\fl d^{(AV)}(x_{12},\, x_{34}) = \frac{1}{M_1^2}\,\frac{C_{10A}(\mu)}{x_{34}}\,\frac{\hat f^{em}_\phi}{x_{34} - \hat M_{2}^2+i\hat \Gamma_{2} \hat M_{2}} \,\Bigg[\frac{ A_{2}(q^2)}{1 + \hat M_{2}  } + \frac{2\hat M_{2}}{x_{12}}\Bigg(A_{3}(q^2) - A_{0}(q^2)\Bigg)\Bigg];
\nonumber
\end{eqnarray}
\\
\begin{eqnarray}
\label{abc_x12x34}
\fl g^{(VA)}(x_{12},\, x_{34}) = \frac{1}{M_1^2}\,\frac{C_{10A}(\mu)}{x_{12}}\,\frac{\hat f^{em}_\phi}{x_{12} - \hat M_{2}^2+i\hat \Gamma_{2} \hat M_{2}} \,\Bigg[\frac{ A_{2}(k^2)}{1 + \hat M_{2}  } + \frac{2\hat M_{2}}{x_{34}}\Bigg(A_{3}(k^2) - A_{0}(k^2)\Bigg)\Bigg];\nonumber
\end{eqnarray}
\\
The dimensionless functions for the bremsstrahlung amplitude (Eq.~\ref{MuBrem}):
\begin{eqnarray}
\fl { d^{(VP)} (x_{12},\, x_{123},\, x_{124})\,}\, =\,{-\,\frac{4 C_{10A} \hat m_e \hat f_{B_s}}{M^2_1}\,
\frac{1}{x_{12}\, (x_{124} - \hat m_e^2)\, (x_{123} - \hat m_e^2)}\,
\frac{(k_3 - k_4,\, q)}{M^2_1}},
\nonumber
\end{eqnarray}
\\
\begin{eqnarray}
\fl {f^{(VT)} (x_{12},\, x_{123},\, x_{124})}\, =\,{-\,\frac{2 C_{10A} \hat m_e \hat f_{B_s}}{M^2_1}\,\frac{1}{x_{12}\, (x_{124} - \hat m_e^2)\, (x_{123} - \hat m_e^2)}\,
\frac{1 + x_{12}-x_{34}}{2}}.
\nonumber
\end{eqnarray}
These are $d$ -- and $f$ -- functions for the ${\mu^+\mu^-}$-- pair emitted by electron and positron in the final state (see first two diagrams on Fig. \ref{fig:Fbrem}). For the ${e^+e^-}$ -- pair emitted by $\mu^+$ and $\mu^-$ functions are similar.\\
In all formulae we use dimensionless variables $x_{12}= q^2/M_1^2$, and $x_{34} =
k^2/M_1^2$, $\hat f^{em}_{\phi} = f^{em}_{\phi}/M_1$, $\hat
M_{2} = M_{2}/M_1$,  and $\hat \Gamma_{2}
= \Gamma_{2}/M_1$. Form factors $F_{TV}(q^2)$, $T_1(q^2)$, $F_V(q^2,k^2)$,
$V(q^2)$, $F_{TA}(q^2,k^2)$, $T_2(q^2)$, $F_A(q^2,k^2)$, $A_1(q^2)$ and $A_2 (k^2)$ are also dimensionless
functions~\cite{MNK}.


\newpage 
\begin{thebibliography}{99}

\bibitem{DarkMatter} M. D. Mauro, M. W. Winkler, "Characteristics of the Galactic Center excess measured with 11 years of Fermi-LAT data", Phys. Rev. D 103, 123005, 2021.
\bibitem{AnMu}B. Abi et al. [Muon g-2 Collaboration], "Measurement of the Positive Muon Anomalous Magnetic Moment to 0.46 ppm", Phys. Rev. Lett. 126, 141801, 2021.
\bibitem{CMS:2014xfa} V.~Khachatryan {\it et al.} [CMS and LHCb Collaborations], ''Observation of the rare $B^0_s\to\mu^+\mu^-$ decay from the combined analysis of CMS and LHCb data'', Nature { 522}, 68, 2015.
\bibitem{Aaboud:2016ire} M.~Aaboud {\it et al.} [ATLAS Collaboration], ''Study of the rare decays of $B^0_s$ and $B^0$ into muon pairs from data collected during the LHC Run 1 with the ATLAS detector'', Eur.\ Phys.\ J.\ C { 76}, no. 9, 513 , 2016.
\bibitem{Aaij:2017vad}    R.~Aaij {\it et al.} [LHCb Collaboration], ''Measurement of the $B^0_s\to\mu^+\mu^-$ branching fraction and effective lifetime and search for $B^0\to\mu^+\mu^-$ decays'', Phys.\ Rev.\ Lett.\  { 118}, no. 19, 191801, 2017.
\bibitem{Aaij:2021}    R.~Aaij {\it et al.} [LHCb Collaboration], ''Measurement
of the $B^0_s\to\mu^+\mu^-$ decay properties
and search for the $B^0\to\mu^+\mu^-$
and $B^0_s\to\mu^+\mu^-\gamma$ decays'', Submitted to Phys.\ Rev.\ D, 2021.
\bibitem{Anom}A.~Dattaa, J.~Kumar, D.~London, "The B anomalies and new physics in $b\to e^+e^-$", Phys.\ Rev.\ B 797, 134858, 2019.
\bibitem{Belle}Belle II experiment: status and prospects, https://aip.scitation.org/doi/abs/10.1063/5.0008685.
\bibitem{Aaij:2016kfs}  R.~Aaij {\it et al.} [LHCb Collaboration], ''Search for decays of neutral beauty mesons into four muons'', JHEP { 1703}, 001, 2017.
\bibitem{Aaij:2013lla}    R.~Aaij {\it et al.} [LHCb Collaboration], ''Search for rare $B^0_{(s)}\rightarrow \mu^+ \mu^- \mu^+ \mu^-$ decays'', Phys.\ Rev.\ Lett.\  { 110}, 211801, 2013.
\bibitem{Aaij:2018pka} R.~Aaij {\it et al.} [LHCb Collaboration], ''Search for the rare decay $B^{+} \rightarrow {\mu}^{+}{\mu}^{-}{\mu}^{+}{\nu}_{{\mu}}$'', Eur.\ Phys.\ J.\ C { 79}, no. 8, 675, 2019.
\bibitem{Dincer:2003zq} Y.~Dincer and L.~M.~Sehgal, ''Electroweak effects in the double Dalitz decay $B_ s \to  \ell^+ \ell^-  \ell'^+ \ell'^-$'', Phys.\ Lett.\ B { 556}, 169, 2003.
\bibitem{Danilina:2018uzr}  A.~V.~Danilina and N.~V.~Nikitin, ''Four-Leptonic Decays of Charged and Neutral $B$ Mesons within the Standard Model'', Phys.\ Atom.\ Nucl.\  {81}, no. 3, 347 (2018) [Yad.\ Fiz.\  {81}, no. 3, 331 (2018)].
\bibitem{BrMu}    A. J. Buras,  M. Munz ''Effective Hamiltonian for B ---> X(s) e+ e- beyond leading logarithms in the NDR and HV schemes'', Phys.\ Rev.\ Lett.\  { 52},  186-195, 1995.
\bibitem{Evt}The development page for the EvtGen project, https://evtgen.hepforge.org/
\bibitem{4mu2021}R.~Aaij {\it et al.} [LHCb Collaboration],"Searches for rare $B^0_s$ and $B^0$ decays into four muons", arXiv:2111.11339.
\bibitem{MNS}D.Melikhov, N.Nikitin, S.Simula, "Rare exclusive semileptonic $b\to s$ transitions in the standard model", 
Phys.\ Rev.\ D~57, p.333, 1998.
\bibitem{MNK} A.~Kozachuk, D.~Melikhov, N.~Nikitin, "Rare FCNC radiative leptonic $B_{d,s}\to \ell^+\,\ell^-\,\gamma$
decays in the standard model", Phys.\ Rev.\ D 97, 053007, 2018.
\bibitem{MN2004}D.~Me\-li\-khov, N.~Nikitin, "Rare radiative leptonic decays $B_{d,s}\to \ell^+\,\ell^-\,\gamma$", Phys.\ Rev.\ D~{70}, 114028, 2004.

\bibitem{Barker:2002ib} A.~R.~Barker, H.~Huang, P.~A.~Toale and J.~Engle, ''Radiative corrections to double Dalitz decays: Effects on invariant mass distributions and angular correlations'',  Phys.\ Rev.\ D { 67}, 033008, 2003.
\end{thebibliography}
\end{document}